\documentclass[11pt,a4paper]{article}
\pdfoutput=1

\usepackage{jcappub}
\makeatletter
\gdef\@fpheader{}
\g@addto@macro\bfseries{\boldmath}
\makeatother

\usepackage{graphicx}
\usepackage{xspace}
\usepackage{epsfig}
\usepackage{subfig}
\usepackage{comment}
\usepackage{slashed}
\usepackage{soul}
\usepackage{tabularx}
\usepackage{physics}
\usepackage[normalem]{ulem}
\usepackage{afterpage}
\usepackage{placeins}
\usepackage{booktabs}
\usepackage{float}
\usepackage{bm}
\usepackage{lipsum} 
\usepackage[]{mdframed}

\usepackage{array}
\newcolumntype{C}[1]{>{\centering\let\newline\\\arraybackslash\hspace{0pt}}m{#1}}

\newcommand{\be}{\begin{equation}} 
\newcommand{\ee}{\end{equation}}
\newcommand{\bea}{\begin{equation}\begin{aligned}} 
\newcommand{\eea}{\end{aligned}\end{equation}}
\newcommand{\ber}{\begin{eqnarray}}
\newcommand{\ear}{\end{eqnarray}}
\newcommand{\Sec}[1]{Sec.~\ref{#1}} 
 
\newcommand{\Fig}[1]{Fig.~\ref{#1}}

\makeatletter
\g@addto@macro\bfseries{\boldmath}
\makeatother

\def\lsim{\mathrel{\raise.3ex\hbox{$<$\kern-.75em\lower1ex\hbox{$\sim$}}}}
\def\gsim{\mathrel{\raise.3ex\hbox{$>$\kern-.75em\lower1ex\hbox{$\sim$}}}}

\renewcommand{\tr}{{\mathrm{Tr}}}

\newcommand{\ie}{{\it i.e.}\xspace}

\newcommand{\td}{{\rm d}}
\newcommand{\pd}{{\partial}}

\newcommand{\Mp}{M_{\scriptscriptstyle{\mathrm{Pl}}}}

\newcommand{\eps}{\epsilon}

\newcommand{\R}{\mathcal{R}}

\newcommand{\I}{\mr{I}}
\newcommand{\II}{\mr{II}}

\newcommand{\mr}[1]{\mathrm{#1}}
\newcommand{\Eq}[1]{Eq.~(\ref{#1})}
\newcommand{\Eqs}[1]{Eqs.~(\ref{#1})}
\newcommand{\T}{T}
\newcommand{\TJ}{T_J}
\renewcommand{\TH}{T_H}

\definecolor{DigitColor}{rgb}{0.5,0.5,0.5}

\definecolor{Red}{rgb}{1,0,0}
\definecolor{Blue}{rgb}{0,0,1}
\definecolor{Green}{rgb}{0,1,0}
\definecolor{Cyan}{rgb}{0,0.8,1}

\newcommand{\kdip}{k_{\mathrm{dip}}}
\newcommand{\kpeak}{{k_{\mathrm{peak}}}}
\newcommand{\knp}{k_{\mathrm{NP}}}
\newcommand{\kone}{k_\I}

\newcommand{\WBessel}[1]{W_{#1}}
\newcommand{\WHankel}[1]{\tilde{W}_{#1}}
\newcommand{\Wsub}{W_{\mathrm{sub}}}
\newcommand{\Wsup}{W_{\mathrm{sup}}}

\begin{document}

\title{How deep is the dip and how tall are the wiggles in inflationary power spectra?}

\author[a]{Vadim Briaud,}
\author[b]{Alexandros Karam,}
\author[b]{Niko Koivunen,}
\author[b,c]{Eemeli Tomberg,}
\author[b]{Hardi Veerm\"{a}e,}
\author[a]{and Vincent Vennin}

\affiliation[a]{Laboratoire de Physique de l'Ecole Normale Sup\'erieure, ENS, CNRS, Universit\'e PSL, Sorbonne Universit\'e, Universit\'e Paris Cit\'e, 75005 Paris, France}

\affiliation[b]{Laboratory of High Energy and Computational Physics, NICPB, \\ R{\"a}vala pst.~10, Tallinn, 10143, Estonia}

\affiliation[c]{Consortium for Fundamental Physics, Physics Department, Lancaster University, \\ Lancaster LA1 4YB, United Kingdom}

\emailAdd{vadim.briaud@phys.ens.fr}
\emailAdd{alexandros.karam@kbfi.ee}
\emailAdd{niko.koivunen@kbfi.ee}
\emailAdd{e.tomberg@lancaster.ac.uk}
\emailAdd{hardi.veermae@cern.ch}
\emailAdd{vincent.vennin@phys.ens.fr}

\abstract{
We study linear scalar perturbations in single-field models of inflation featuring a non-attractor phase. These models lead to a peak in the curvature power spectrum that may result in the formation of primordial black holes. We develop a transfer-matrix formalism, analogous to the S-matrix program in quantum-field theory, that maps perturbations throughout the transitory phase. At scales smaller than the peak, the power spectrum features damped oscillations, and the duration of the transition sets the scale at which power-law damping switches to exponential damping. At scales larger than the peak, we demonstrate that a dip appears in the power spectrum if and only if the inflaton's velocity does not flip sign. We show that the amplitude at the dip always scales as the inverse square-rooted amplitude of the peak, and comment on the physical consequences of this universal relationship. We also test the robustness of our results with a few toy models and interpret them with an intuitive mechanical analogy.
}

\maketitle

\section{Introduction}
\label{sec:intro}

Inflation is the leading paradigm to describe the early universe, addressing fundamental issues in the standard Big Bang model, such as the horizon and flatness problems~\cite{Starobinsky:1980te, Kazanas:1980tx, Sato:1980yn, Guth:1980zm, Linde:1981mu, Albrecht:1982wi, Linde:1983gd}. Beyond resolving these problems, inflation offers a natural mechanism for generating primordial quantum fluctuations that are stretched to cosmological scales and later evolve into the observed large-scale structure (LSS) of the universe and the anisotropies in the cosmic microwave background (CMB)~\cite{Starobinsky:1979ty, Mukhanov:1981xt, Hawking:1982cz, Starobinsky:1982ee, Guth:1982ec, Bardeen:1983qw}. At small scales, the lack of constraints from either CMB or LSS opens the possibility for significant enhancements in the scalar power spectrum, which could lead to the formation of primordial black holes (PBHs)~\cite{Hawking:1971ei, Carr:1974nx, Carr:1975qj}. PBHs are intriguing candidates for dark matter~\cite{Niikura:2017zjd, Katz:2018zrn, Montero-Camacho:2019jte} and may also play a role in the formation of supermassive black holes observed at high redshifts~\cite{1975A&A....38....5M, Duechting:2004dk, Kawasaki:2012kn, Clesse:2015wea, Carr:2018rid, Liu:2022bvr, Hutsi:2022fzw}.

The evolution of scalar perturbations during inflation is governed by the Mukhanov--Sasaki equation~\cite{Mukhanov:1981xt, Sasaki:1986hm}, which describes the dynamics of the curvature perturbation $\mathcal{R}$ in a perturbed homogeneous and isotropic spacetime. The power spectrum $\mathcal{P}_\mathcal{R}(k)$, which quantifies the variance of these perturbations as a function of scale $k$, is one of the key observables linking the inflationary potential to the distribution of PBHs. The power spectrum is nearly scale-invariant in standard slow-roll (SR) inflationary models. However, in scenarios featuring phases where slow roll is violated, such as ultra-slow roll (USR)~\cite{Dimopoulos:2017ged}, the power spectrum can exhibit non-trivial features such as dips, peaks and oscillations. The peaks correspond to enhanced perturbations at small scales and are thus crucial for PBH formation. 

In this work, we consider models whose inflationary timeline follows the typical phases seen in single-field PBH scenarios: the scales observed in the CMB are generated during an initial SR phase, which then transitions into a USR phase, causing the inflaton to slow down significantly. This is followed by another phase of SR or constant-roll (CR) inflation~\cite{Karam:2022nym}. The models of this type considered in the literature can be classified into four categories: {\it i)} Quasi-inflection point potentials~\cite{Garcia-Bellido:2017mdw, Kannike:2017bxn, Ballesteros:2017fsr, Germani:2017bcs, Motohashi:2017kbs, Ezquiaga:2017fvi, Di:2017ndc, Hertzberg:2017dkh, Rasanen:2018fom, Cicoli:2018asa, Ozsoy:2018flq, Gao:2018pvq, Atal:2019cdz, Atal:2019erb, Mishra:2019pzq, Ballesteros:2019hus, Dalianis:2018frf, Bhaumik:2019tvl, Drees:2019xpp, Dalianis:2019asr, Ballesteros:2020qam, Ragavendra:2020sop, Nanopoulos:2020nnh, Iacconi:2021ltm, Stamou:2021qdk, Wu:2021zta, Ng:2021hll, Rezazadeh:2021clf, Wang:2021kbh, Gu:2022pbo, Frolovsky:2022qpg, Cicoli:2022sih, Ghoshal:2023wri,Pi:2022zxs,Allegrini:2024ooy}; {\it ii)} potentials with an upward or downward step~\cite{Cai:2021zsp, Inomata:2021tpx, Inomata:2021uqj, Kefala:2020xsx, Dalianis:2021iig}; {\it iii)} models where the inflaton rolls through a global minimum/double-well potentials~\cite{Yokoyama:1998pt, Saito:2008em, Bugaev:2008bi, Fu:2020lob, Briaud:2023eae, Karam:2023haj}; and {\it iv)} potentials with multiple features or oscillations~\cite{Cai:2019bmk, Tasinato:2020vdk, Zhou:2020kkf, Peng:2021zon, Inomata:2022yte, Fumagalli:2023loc, Caravano:2024tlp}.

The amplification of the power spectrum depends on the way these transitions occur. For instance, in double-well potentials~\cite{Briaud:2023eae, Karam:2023haj} the field traverses a global minimum during the transition from SR to USR. The characteristics of the peak are mainly shaped by the details of that transition, which results in pronounced oscillatory spectral features, similar to those arising from particle production during preheating.
In contrast, in quasi-inflection point scenarios~\cite{Garcia-Bellido:2017mdw, Germani:2017bcs, Ballesteros:2017fsr} (see~\cite{Karam:2022nym, Ozsoy:2023ryl} for recent reviews), the enhancement of the power spectrum is largely determined by the duration of the USR phase and the Wands duality~\cite{Wands:1998yp}, which relates the USR phase to the subsequent CR phase. 

The behaviour of the power spectrum during the rapid transition from SR to non-attractor inflation is crucial for understanding PBH phenomenology. Therefore, developing analytic models to describe this behaviour is highly valuable. This can be achieved by considering toy models where the evolution of the Hubble-flow parameters is assumed rather than derived from solving the field equations. For instance, in~\cite{Byrnes:2018txb}, it was demonstrated that in single-field inflation, regardless of the shape of the inflaton potential, the steepest possible growth of the primordial power spectrum is $\mathcal{P}_\mathcal{R} \propto k^4$. A more detailed investigation~\cite{Carrilho:2019oqg} demonstrated that an even steeper growth, characterised by $\mathcal{P}_\mathcal{R} \propto k^5 \left( \ln k \right)^2$, is possible (see also~\cite{Cai:2019bmk, Liu:2020oqe, Ozsoy:2019lyy, Ballesteros:2020sre, Ragavendra:2020sop, Tasinato:2020vdk, Ng:2021hll, Ozsoy:2021pws, Inomata:2024dbr, Cielo:2024poz}). However, such growth beyond $k^4$ has limited effect on the PBH mass function~\cite{Cole:2022xqc} and necessitates an initial phase of inflation with a blue-tilted spectrum~\cite{Karam:2022nym}.

The Mukhanov--Sasaki equation admits two solutions, a growing mode and a decaying mode. In CR, the growing mode dominates over the decaying one. At super--Hubble scales, the curvature perturbation can be tracked by following the dominant mode only and this can be achieved using non-perturbative techniques such as the separate-universe or $\delta N$ formalisms~\cite{Starobinsky:1982ee, Starobinsky:1985ibc, Sasaki:1995aw, Sasaki:1998ug, Lyth:2004gb}. However, during the transition, the two modes become comparable at mildly super--Hubble scales, and these approaches cannot be employed directly~\cite{Jackson:2023obv, Artigas:2024xhc, Jackson:2024aoo}. 

In this paper, we study the dynamics of inflation perturbations at linear order, in models featuring transitions between attractor and non-attractor phases. Our goal is to determine under which conditions a dip appears in the power spectrum, how its depth is related to the amplitude of the peak, and to characterise the presence and decay of oscillatory features at scales smaller than the peak. We do so by developing a transfer-matrix formalism, which allows for controlled expansions in different regimes.

The layout of the paper is as follows. In the next section, \Sec{sec:inflationary_perturbations}, we define the main phases of PBH single-field models, and qualitatively discuss the behaviour of scalar perturbations in terms of a mechanical analogy. 
Then, in \Sec{sec:transfermarix} we develop a transfer-matrix approach and perform a quantitative study of the transition and of its effect on the perturbations. In \Sec{sec:subhorizon}, we apply this formalism to describe oscillations at the peak of the power spectrum. In \Sec{sec:suphorizon}, we derive a criterion for the existence of the dip (namely a dip exists if and only if the inflaton's velocity does not flip sign), and relate its amplitude to the height of the peak. The transfer-matrix formalism is then applied to the cases of an instantaneous and a smooth transition in \Sec{sec:instantaneous_transitions} and \Sec{sec:example_damping}, respectively. Finally, we conclude in \Sec{sec:concl}. Some computational details are deferred to the appendices, and natural units $c = \hbar = 1$ are used throughout this paper.

\section{Inflationary perturbations in single-field PBH models}
\label{sec:inflationary_perturbations}

If inflation is driven by a single scalar field $\phi$, the scalar sector of cosmological perturbations reduces to a single physical degree of freedom, that can be described by the curvature perturbation $\zeta$. At linear order, its dynamics is determined by the action~\cite{Mukhanov:1981xt, Sasaki:1986hm}
\bea
    S=\frac{1}{2} \int\dd^4x \left[(u')^2-\delta^{ij}\partial_i u \partial_j u+\frac{z''}{z}u^2\right] ,
\eea 
where $u\equiv z\zeta$ is the so-called  Mukhanov--Sasaki variable, and $z \equiv a \dot\phi/H$ where $a$ is the scale factor and $H$ is the Hubble parameter of the background Friedmann-Lema\^itre-Robertson-Walker metric. A dot denotes derivation with respect to the cosmic time $t$. In Fourier space, the Euler-Lagrange equations give rise to
\be\label{eq:MS}
   u''_k + \left( k^2 - \frac{z''}{z} \right) u_k = 0\, ,
\ee
where a prime denotes derivation with respect to conformal time $\tau=\int\dd t/a$.

In what follows, it is convenient to introduce the Hubble-flow parameters $\epsilon_i$, defined as $\epsilon_1 \equiv -\dot{H}/H^2$ and $\epsilon_{i+1} \equiv \td\ln \epsilon_i/\td N$ where $N=\ln(a)$ is the number of e-folds. Inflation, defined as $\ddot{a}>0$, takes place when $\epsilon_1<1$. In terms of the Hubble-flow parameters, the time-dependent part of the frequency appearing in the Mukhanov--Sasaki equation reads
\bea
\label{eq:nu2:def}
    \frac{z''}{z}=(aH)^2\left(\nu^2-\frac{1}{4}\right)
    \quad\text{where}\quad
    \nu^2=\frac{9}{4}-\epsilon_1+\frac{3}{2}\epsilon_2-\frac{1}{2}\epsilon_1\epsilon_2+\frac{1}{4}\epsilon_2^2+\frac{1}{2}\epsilon_2\epsilon_3\, ,
\eea
which also defines $\nu^2$. In SR inflation, all Hubble-flow parameters are small, hence $\nu^2\simeq 9/4$. In USR inflation, all Hubble-flow parameters are small except $\epsilon_2\simeq -6$, which also leads to $\nu^2\simeq 9/4$. SR and USR inflation are thus related by the so-called Wands duality~\cite{Wands:1998yp}. 

\subsection{Constant-$\nu$ inflation}
\label{sec:Constant:Roll}

More generally, if the potential energy stored in the inflaton, $V(\phi)$, is a sufficiently smooth function, in most cases inflation proceeds in the constant-$\nu$ regime, where $\nu^2$ varies slowly. This is the case in SR and USR, but also in ``uphill inflation''~\cite{Briaud:2023eae}, where the inflaton climbs up its potential close to a local maximum and $\nu^2$ asymptotes a constant value.

Indeed, in models giving rise to large peaks in the power spectrum, the inflaton must slow down significantly, hence it does not vary much during the phase where the peak is generated. The potential function can then be expanded around a local maximum $\phi_r$, and the Klein-Gordon equation $\ddot \phi + 3H \dot \phi + V' = 0$ becomes~\cite{Karam:2022nym}
\be
\label{eq:eom_hilltop}
    \pd^2_N \phi + 3 \pd_N \phi + 3\eta_V(\phi_r) (\phi - \phi_r) = \mathcal{O}(\phi - \phi_r)^2\, ,
\ee
where $\eta_V \equiv \Mp^2 V''/V$ is the second potential slow-roll parameter. Note that $\eta_V(\phi_r)<0$, and since $\dot\phi^2 \ll V$ during this epoch, we used $3\Mp^2 H^2 \simeq V$ for the Friedmann equation. Neglecting higher-order terms in $\phi - \phi_r$, we find that the inflation evolves as
\be
\label{eq:phi_sol_hilltop}
    \phi - \phi_r 
    = \phi_+ a^{-3/2+\nu_V} + \phi_- a^{-3/2 - \nu_V}\,
    \qquad \mbox{where} \qquad
    \nu_V = \frac{3}{2}\sqrt{1-\frac{4}{3}\eta_V(\phi_r)}\,.
\ee
The first term describes an accelerating field with $\epsilon_{2+} \geq 0$ and produces a constant roll (CR) phase when dominant. The second term describes a decelerating field with $\epsilon_{2-} \leq -6$ and corresponds to a USR-like epoch. These phases satisfy the Wands dual relation $\epsilon_{2+} + \epsilon_{2-} = -6$.

\begin{figure}[t]
\centering
\includegraphics[height=0.33\textwidth,trim={0cm -0.8cm 0cm 0cm}, clip]{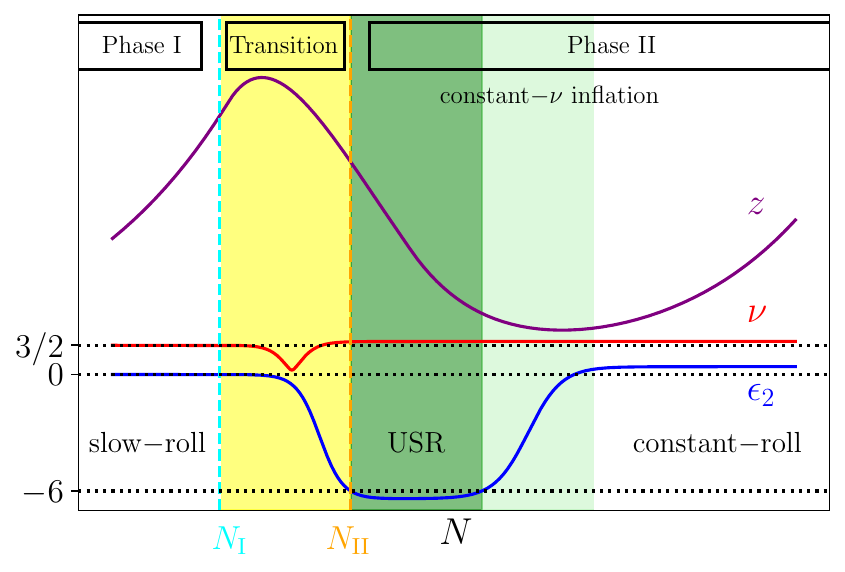}
\includegraphics[height=0.33\textwidth,trim={0cm 0cm 0cm 0cm}, clip]{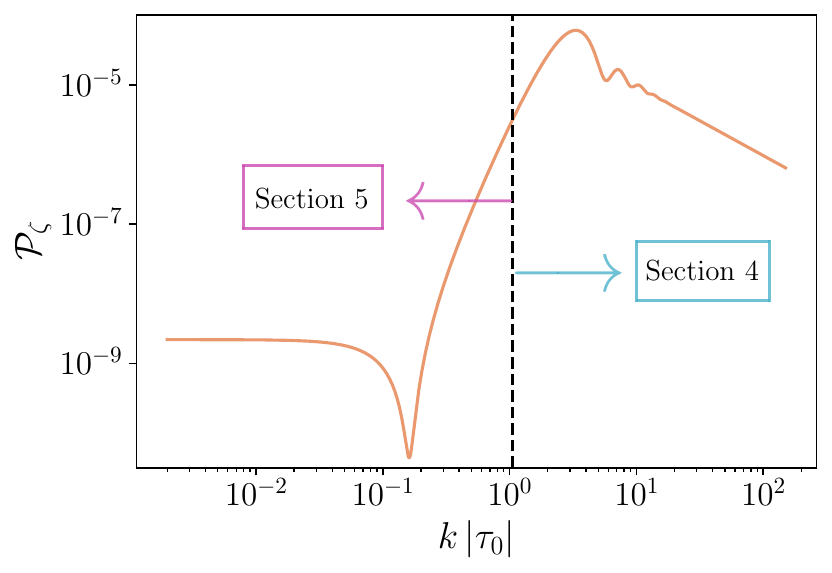}
\caption{\emph{Left panel:} Schematic evolution of the cosmological parameters $\epsilon_2$ (in blue), $\nu$ (in red), and $z$ (in purple) with the number of $e$-folds during the transition from an initial slow-roll phase to a final constant-roll phase. We display remarkable values of those parameters in black dotted lines, $\epsilon_2 = -6$ corresponding to an exact USR phase, $\epsilon_2 \approx 0$ corresponding to an SR phase, and $\nu = 3/2$ describing both an SR or a USR phase. The e-folding at the beginning and at the end of the SR-USR transition is also displayed with vertical lines respectively in cyan and orange. \emph{Right panel:} Schematic power spectrum required to produce PBH. The spectrum features a dip at large scales and damped oscillations that follow a peak at small scales.}
\label{fig:schematic plot}
\end{figure}

At this stage, the lexical convention adopted in this work is worth clarifying: ``constant roll'' refers to a phase where $\epsilon_2$ is a constant, whereas ``constant $\nu$'' is devoted to regimes where $\nu$ is constant. The dynamics described by \Eq{eq:phi_sol_hilltop} proceeds at constant $\nu$ all along, and it is made of two (distinct) constant-roll phases, during which a different branch dominates. This can be seen in the sketch drawn in \Fig{fig:schematic plot}: after the transition (labeled in yellow), $\nu$ remains constant, while $\epsilon_2$ goes from one constant to the other, hence two constant-roll regimes follow each other. 

Let us note that the presence of a local maximum is not a strong restriction on the argument. The absence of a local maximum would introduce a constant term $\propto V'(\phi_r)$, but since the potential must be extremely flat for the generation of the spectral peak, such a term can often be neglected. This is why the CR and USR phases that follow from \Eq{eq:phi_sol_hilltop} are realised in all quasi-inflection point models for PBHs in practice (these models are often dubbed ``USR models'' even if an exact USR phase with $\epsilon_2 = -6$ is not realised).

The background and the perturbations follow a very generic behaviour in this setup. Indeed, in single-field inflation, the time-dependent mass term in the Mukhanov--Sasaki equation~\eqref{eq:nu2:def} can be rewritten as
\bea
    \frac{z''}{z} 
    &= (aH)^2\left[2 - (3-\eps_1) \eta_V + \eps_1 (5-2\eps_1+2\eps_2)\right] 
    = (aH)^2\left(\nu_V^2-\frac{1}{4}\right) + \mathcal{O}(\eps_1)\, ,
\eea
implying that
\be
    \nu \simeq \nu_V
\ee
is practically constant as the inflaton transits from the USR phase to its dual CR phase in a way described by \Eq{eq:phi_sol_hilltop}. Consequently, the behaviour of $z''/z$ does not change much during this transition, hence the evolution of the $u_k$ modes is mostly insensitive to it.  
Nevertheless, this does not hold for the curvature perturbation $\zeta_k = u_k/z$ because $z$ evolves differently in the two phases. In particular, the rapid damping of $z$ during the USR-like phase is responsible for the enhancement of $\R_k$.

The Mukhanov--Sasaki equation, being second order, admits two independent solutions usually referred to as two ``modes''. At super-Hubble scales, \ie~when $k\ll a H$, a clear hierarchy develops between the two modes in constant-$\nu$ regimes. 
This implies that the curvature perturbation can be tracked by following the dominant mode only. This can be done also at the non-perturbative level using techniques such as the separate-universe or $\delta N$ formalisms~\cite{Starobinsky:1982ee, Starobinsky:1985ibc, Sasaki:1995aw, Sasaki:1998ug, Lyth:2004gb}.

When two phases of constant-$\nu$ inflation follow each other, the required presence of features in the inflationary potential implies that a transitory phase takes place, during which the two modes become comparable at mildly super-Hubble scales. This means that the above-mentioned techniques cannot be applied directly~\cite{Jackson:2023obv, Artigas:2024xhc, Jackson:2024aoo}. These scenarios have nonetheless received increasing attention since they can amplify cosmological perturbations at small scales, and may thus induce the formation of primordial black holes~\cite{Byrnes:2018txb, Cole:2022xqc, Karam:2022nym}. 

This is why in this work, we will consider the generic scenario sketched in~\Fig{fig:schematic plot}, where constant-roll inflation is interrupted by an intermediate transitory phase that shifts the system to the ultra-slow-roll dynamics before relaxing to another constant-roll regime. The inflationary timeline can then be divided into the following phases~\cite{Karam:2022nym}:
\begin{description}
    \item[($\I$)] An initial constant-roll attractor phase,
    \item[($\mr{T}$)] A transitory phase that interrupts constant roll or even ends inflation,
    \item[($\II$)] A constant-$\nu$ phase, which can be divided into three sub-phases :
\begin{description}
    \item[($\II_{\mr{USR}}$)] A USR-like transitory phase responsible for slowing down the inflation,
    \item[($\II_{\mr{T}}$)] A transition from USR to CR,
    \item[($\II_{\mr{CR}}$)] A second constant-roll attractor phase.
\end{description}
\end{description}
In most models studied in the literature, phase $\I$ proceeds in the slow-roll regime, but we allow for attractor constant roll with any $\epsilon_2 > -3$.\footnote{If $\phi\propto a^p$, then $\epsilon_2 = 2p$. Therefore, when the first branch dominates at late time in \Eq{eq:phi_sol_hilltop}, one has $\epsilon_2=-3+2\nu>-3$, and $\epsilon_2<-3$ when the second branch still dominates. An attractor behaviour thus requires $\epsilon_2>-3$.}  
Phase $\II$ proceeds at the same constant $\nu$ and can be described by the above formulas.
Therefore, the only non-trivial phase is the transition $\mr{T}$, to which we shall now turn our attention. The transition causes the spectral features seen in~\Fig{fig:schematic plot}: a dip in power, followed by steep growth, which transitions into damped oscillations. All in all, while the USR to SR (or CR) transition is trivially accounted for in phase II due to $\nu$ being constant, the SR to USR transition and its effect on the spectrum can be highly non-trivial and has to be studied in detail.

\subsection{A mechanical analogy}
\label{sec:Modulus:MSvariable}

Before studying the spectral features quantitatively, some intuition can be built by unpacking the Mukhanov--Sasaki variable $u_k$ into its modulus and argument:
\begin{equation} \label{eq:u_polar}
    u_k = r_k e^{i\theta_k} \, .
\end{equation}
Then, the real and imaginary parts of the Mukhanov--Sasaki equation \eqref{eq:MS} read
\begin{equation} \label{eq:r_theta_eom}
    r_k'' + \qty(k^2 - \frac{z''}{z} - \theta_k'^2)r_k = 0 \, , \qquad
    2r_k'\theta_k' + r_k\theta_k'' = 0 \, .
\end{equation}
The second equation can be integrated to yield $r_k^2 \theta_k' = L$, where the integration constant $L$ is analogous to the conserved angular momentum in a rotationally-invariant mechanical system.\footnote{The conservation of the angular momentum is analogous to the conservation of the Wronskian $W(u_k, u_k^*) = u_k {u_k^*}' - u_k^* u_k' = -2i\theta_k' r_k^2 = -2iL = i$.} In this mechanical analogy, the spatial coordinates of a test particle are replaced by the real and imaginary parts of $u_k$. The Bunch--Davies initial condition at $\tau\to -\infty$, $u_k \sim e^{-i k\tau}/\sqrt{2k}$, gives $r_k \sim 1/\sqrt{2k}$, $r'_k \sim 0$ and $\theta_k \sim -k\tau$. This fixes $L=-1/2$, so that $\theta_k' = -1/(2r_k^2)$. Plugging everything back into the equation of motion \Eq{eq:r_theta_eom}, we obtain
\begin{equation} \label{eq:r_V_eom}
    r_k'' +  V_k'(r_k) = 0 \, ,
\end{equation}
where we defined the effective potential
\begin{equation} \label{eq:Vk}
    V_k(r_k) = \frac{1}{2}\qty(k^2 - \frac{z''}{z})r_k^2 + \frac{1}{8r_k^2} \, 
\end{equation}
to reinforce the mechanical analogy.

Note that $V_k$ has explicit time dependence through $z$, so there is no energy conservation -- only the angular momentum is conserved. It also diverges when $r_k \to 0$, which is due to the last term in \eqref{eq:Vk}, arising from the non-vanishing angular momentum. This protects the mode from ever reaching the origin and implies that the power spectrum can never vanish exactly, even at the dip.

\subsubsection{Oscillations at small scales}
\label{sec:mechanical:analogy:oscillations}

At sub-Hubble scales, $z''/z$ is subdominant in the effective potential $V_k$, which thus features a minimum at $r_k=1/\sqrt{2k}$, the Bunch--Davies vacuum. The modulus $r_k$ starts at rest there, while the argument $\theta_k$ rotates around the origin.

Let us consider a mode $k$ that is still sub-Hubble at the transition. It sits at the potential minimum until the transition comes with a brief pulse in $z''/z$ that displaces the minimum before bringing it back to its initial location. If the transition is slow, then $r_k$ adiabatically follows the minimum as it moves around, but if the transition is sharp then $r_k$ is shifted away from the minimum and starts oscillating. The potential's second derivative at the minimum is $V''(1/\sqrt{2k}) = 4k^2$, leading to oscillations with an angular frequency of $2k$,
\begin{equation}\label{eq:r_k_oscillations}
    r_k - \frac{1}{\sqrt{2k}} \propto \sin\left[2k(\tau - \tau_0) + \varphi_0\right] .
\end{equation}
Here, $\tau_0$ is the moment of the transition and $\varphi_0$ is an approximately $k$-independent phase induced by the transition.
Later, when the mode exits the Hubble radius around $k\tau = -1$, $z''/z$ overtakes the other terms in $V_k$, which becomes unstable. This makes $r_k$ grow, ultimately like $r_k \propto z$, and the curvature perturbation $\zeta_k \propto r_k/z$ freezes.

When the instability starts at $\tau=-1/k$, oscillations stop and the phase in \Eq{eq:r_k_oscillations} remains fixed. Up to a constant shift, that phase grows linearly with $k$ as $-2k\tau_0$ so we expect oscillations with angular frequency $2\tau_0$ in the power spectrum. The disruption from the pulse is smaller at larger values of $k$, hence the oscillations are damped at small scales, see \Fig{fig:schematic plot}. They are studied in more detail in \Sec{sec:subhorizon}.

\subsubsection{Dip at large scales}
\label{sec:dip_or_no_dip}
Let us next consider modes that are super-Hubble at the transition $\mr{T}$, for which the $z''/z$ term in \eqref{eq:Vk} dominates. In phase $\I$, $r_k$ follows the growing mode $z$. As depicted in \Fig{fig:schematic plot}, $z$ first grows, then decreases after the transition, and finally grows again. Below, we will also consider cases where $z$ crosses zero during phase $\II_{\mr{T}}$, so its final growth takes place with an opposite sign, see \cite{Karam:2022nym}. This corresponds to the inflaton field coming to a halt and rolling back.

Since the effective potential $V_k$ diverges at the origin, $r_k$ cannot be brought arbitrarily close to zero, which implies that it necessarily stops tracking $z$ in case $z$ flips sign. This entails that gradient contributions become relevant at some point during the transition, which induces extra $k$-dependence in the power spectrum. Even if $z$ does not flip sign, $z''/z$ becomes subdomiant  around the inflection point of $z$ during the transition; nonetheless, due to the low-$r_k$ behaviour, different features should be expected in the power spectrum depending on whether $z$ flips sign or not. Below, we will show that this is indeed the case: if $z$ does not flip sign, there is a dip in the power spectrum, but if a sign flip happens, there is no dip. We will study the dip, or lack thereof, in more detail in \Sec{sec:suphorizon}.

\section{Transfer-matrix approach to inflationary features}
\label{sec:transfermarix}

Let us now turn to a quantitative study of the transition T and of its effect on the perturbations. This effect is known analytically only in limiting cases. To allow for a systematic expansion around these limits, we propose to describe transitory phases with a transfer matrix $\T$, so that 
\be\label{eq:MS_phases_tildeT}
    \begin{pmatrix}
    u_k(\tau_\II) \\ u'_k(\tau_\II)
    \end{pmatrix}
    =
    \T(k)
    \begin{pmatrix}
    u_k(\tau_\I) \\ u'_k(\tau_\I)
    \end{pmatrix}\, ,
\ee
where $\tau_1$ and $\tau_2$ are the times the transitory phase starts and ends. In this work, we study the properties of $\T$ and find methods to approximate it in different regimes. At linear order in perturbation theory, it fully characterises the transitory phase and its effect on the power spectrum of the curvature perturbation. To construct it explicitly, let us introduce first the Wronskian matrix of two functions $u_1(\tau)$ and $u_2(\tau)$, defined as
\bea\label{eq:Wronskian}
    W\left(u_1,u_2\right) \equiv
    \begin{pmatrix}
    u_1 & u_2\\
    u_1' & u_2'
    \end{pmatrix}\, .
\eea
Now, given two independent solutions of the Mukhanov--Sasaki equation \eqref{eq:MS}, $u_{1,k}(\tau)$ and $u_{2,k}(\tau)$, Eq.~ \eqref{eq:MS_phases_tildeT} can be recast as $W\left[u_{1,k}(\tau_{\II}),u_{2,k}(\tau_{\II})\right] = \T(k)  W\left[u_{1,k}(\tau_{\I}),u_{2,k}(\tau_{\I})\right]$, so the transfer matrix reads
\be \label{eq:T:in:W:and:W-1}
    \T(k) = W\left[u_{1,k}(\tau_{\II}),u_{2,k}(\tau_{\II})\right] W\left[u_{1,k}(\tau_{\I}),u_{2,k}(\tau_{\I})\right]^{-1}\,.
\ee

A priori, the transfer matrix is a $2\times 2$ complex matrix containing $8$ unknown real coefficients. However, it must satisfy two constraints. First, the Mukhanov--Sasaki equation~\eqref{eq:MS} being real, it is always possible to pick two real solutions $u_{1,k}$, $u_{2,k}$. Since the transfer matrix $\T(k)$ does not depend on this choice, it must be real as well. Second, the dynamics of cosmological perturbations is unitary, which in practice implies that the Wronskian is conserved. 
This can be checked directly by differentiating $\det [W(u_1,u_2)]=u_1 u_2' - u_1' u_2$ with respect to $\tau$ and assuming $u_1$ and $u_2$ obey the Mukhanov--Sasaki equation \eqref{eq:MS}. As a result,
\be
    \det[\T(k)] = 1\,,
\ee
since $\det[W\left(u_{1,k}(\tau_{\II}),u_{2,k}(\tau_{\II})\right)] = \det[W\left(u_{1,k}(\tau_{\I}),u_{2,k}(\tau_{\I})\right)]$ due to the conservation of the Wronskian.

Finally, $\T(k)$ has a symplectic structure. To show this explicitly, let us adopt a vectorial notation where 
\bea
\label{eq:U:def}
\bm{U}(\tau)=
\begin{pmatrix}
u(\tau)\\
u'(\tau)
\end{pmatrix}\, ,
\eea
hence \Eq{eq:MS_phases_tildeT} can be rewritten as $\bm{U}(\tau_\II)=\T\bm{U}(\tau_\I)$. One has
\bea
    \det\left[ W\left(u_1,u_2\right)\right] = \bm{U}_2^{\mathrm{T}} \Omega \bm{U}_1,
\qquad\text{where}\qquad
\Omega \equiv 
\begin{pmatrix}
0 & -1\\
1 & 0
\end{pmatrix}\, .
\eea
Together with \Eq{eq:MS_phases_tildeT}, one can write the determinant of the Wronskian at time $\tau_\II$ as $\det W=\bm{U}_2^{\mathrm{T}}(\tau_\II) \Omega \bm{U}_1(\tau_\II) =\bm{U}_2^{\mathrm{T}}(\tau_\I) \T^{\mathrm{T}} \Omega \T \bm{U}_1(\tau_\I) $. This must be equal to the Wronskian at time $\tau_\I$, namely $\det W=\bm{U}_2^{\mathrm{T}}(\tau_\I)  \Omega   \bm{U}_1(\tau_\I) $, for all vectors $\bm{U}_1(\tau_\I) $ and $\bm{U}_2(\tau_\I) $. One concludes that
\bea
\label{eq:symplectic:relation}
 \T^{\mathrm{T}} \Omega \T =  \Omega \, .
\eea
This relation is the one defining the symplectic group, so $\T$ is a real symplectic matrix. It implies that unitary evolution can be seen as a canonical transformation~\cite{Grain:2019vnq}, and although we focused on single-field models where there is a single Mukhanov--Sasaki variable, the result would hold in higher dimensions, \ie for multi-field models. In dimension $2$, the real symplectic group $\mathrm{Sp}(2,\mathbb{R})$ is made of all real matrices with unit determinant, $\det \T=1$, as can be explicitly checked from \Eq{eq:symplectic:relation}.\footnote{This can also be shown by noticing that \Eq{eq:MS_phases_tildeT} implies that $W(\tau_\II)=\T W(\tau_\I)$, hence for $W(\tau_\I)$ and $W(\tau_\II)$ to have the same determinant one must have $\det \T=1$.\label{footnote:W=TW}} The transfer matrix thus contains three real coefficients only (in the Bloch-Messiah decomposition~\cite{Bloch:1962zj}, real symplectic matrices are parametrised by two angles and one squeezing amplitude). When expanding $T$ around a certain limit, this should thus be done in the Lie algebra of $\mathrm{Sp}(2,\mathbb{R})$. 

The eigenvalues of $\T$ are solutions to its characteristic equation $ \lambda^2 - \tr (\T) \lambda + 1 = 0$, and are thus determined only by $\tr ( \T)$. Importantly, $\T$ has real eigenvalues when
\be
\label{eq:Tr:gt:2}
    |\tr ( \T)| \geq 2 \, .
\ee
Recall that $\T$ is always a real matrix, but its eigenvalues are real only if it can be diagonalised on $\mathbb{R}$, and this is the case only if \Eq{eq:Tr:gt:2} is satisfied.

Regarding the $k$-dependence of $T$, we note that since the function $z(\tau)$ obeys the Mukhanov--Sasaki equation~\eqref{eq:MS} for $k=0$, we have
\be\label{eq:MS_z_tildeT}
    \begin{pmatrix}
    z(\tau_\II) \\ z'(\tau_\II)
    \end{pmatrix}
    =
    \T(0)
    \begin{pmatrix}
    z(\tau_\I) \\ z'(\tau_\I)
    \end{pmatrix}\,.
\ee

\subsection{Connecting two constant-$\nu$ phases}
\label{sec:connect:two:constant:nu}

Let us now consider the situation of \Sec{sec:Constant:Roll}, where the transitory phase connects two constant-$\nu$ phases. We denote by $\nu_\I$ and $\nu_\II$ the (quasi-constant) value of the $\nu$ parameter in these two phases respectively. We also assume that $\epsilon_1\ll 1$ in both constant-$\nu$ phases, hence $aH\simeq -1/\tau$ and the solutions to the Mukhanov--Sasaki equation~\eqref{eq:MS} can be expanded onto Bessel functions,
\bea
\label{eq:expansion:Bessel}
     u_{i,k}(\tau)
     =  a_{i,1}(k) u^{(1)}_i(k,\tau)  + a_{i,2}(k) u^{(2)}_i(k,\tau) \,,
\eea
where $i \in \{\I, \II\}$ labels the two constant-$\nu$ phases and we defined the mode functions
\bea
\label{eq:modes_bessel}
    u^{(1)}_i(k,\tau) 
    & =  \frac{\sqrt{-\pi\tau}}{2\sin(\pi\nu_i)}J_{-\nu_i}\left(-k\tau\right) \, ,
    \qquad
   u^{(2)}_i(k,\tau) 
    =  -\sqrt{-\pi\tau}J_{\nu_i}\left(-k\tau\right)  \, .
\eea
They are normalised so that their Wronskian determinant is to unity. By convention, we choose $\Re(\nu)>0$, which is always possible since only $\nu^2$ is defined in \Eq{eq:nu2:def}. The $a_{i,1}$ and $a_{i,2}$ coefficients depend on the initial conditions. Since $J_\nu(x)\propto x^\nu$ when $x\ll 1$, they control the growing and decaying modes, respectively. At late time, the growing mode dominates, which is why the amplitude of late-time perturbations is determined by $a_{\II,1}$. We assume the first constant-$\nu$ phase to last long enough such that, for the scales of interest, the Bunch--Davies vacuum can be used as an initial condition $u_{\I,k}(\tau\to -\infty) \sim e^{-i k\tau}/\sqrt{2k}$, yielding
\be
\label{eq:a:BD}
    \bm{a}_{\I}(k) = e^{\frac{i\pi}{2}\left(\nu_\I-\frac12\right)}
    \begin{pmatrix}
    1 \\   \frac{e^{-i\pi \nu_\I}}{2\sin(\pi\nu_\I)}
    \end{pmatrix}\, ,
\ee
where we have introduced the vector $\bm{a}_{i} \equiv (a_{i,1},a_{i,2})$. 

Instead of mapping $u$ and $u'$ at both ends of the transitory phase through the transfer matrix $\T$, another way to describe the transition is to relate the growing- and decaying-mode coefficients,
\be\label{eq:MS_phases_T}
     \bm{a}_\II(k)
    = \TJ (k)  \bm{a}_\I(k)\, .
\ee
This defines the matrix $\TJ$, which can be seen as a transfer matrix in a different representation. Rewriting \Eq{eq:expansion:Bessel} in a matricial form as $\bm{U}_i = W(u^{(1)}, u^{(2)}) \bm{a}_i $, where $W$ is the Wronskian \eqref{eq:Wronskian}, \Eqs{eq:MS_phases_tildeT} and~\eqref{eq:MS_phases_T} allow $\TJ$ to be related to $\T$ via
\bea
\label{eq:TJ}
    \TJ (k) = \WBessel{\II}^{-1}(k)\T (k) \WBessel{\I}(k)\, ,
\eea
where we use the shorthand notation
\bea
\label{eq:W_Bessel}
    \WBessel{i} = W\left(u^{(1)}_i,u^{(2)}_i\right) ,
\eea
with $u^{(1)}_i$ and $u^{(2)}_i$ the Bessel-basis functions given in \Eq{eq:modes_bessel}. These functions were normalised in such a way that their Wronskian has unit determinant, $\det \WBessel{i}=1$. This is why \Eq{eq:TJ} implies that $\TJ$ also has unit determinant. Moreover, the functions $u^{(1)}_i$ and $u^{(2)}_i$ have also been chosen as real, hence the $\WBessel{i}$ matrices are real and it follows from \Eq{eq:TJ} that  $\TJ$ is real. As a consequence, $\TJ$ is a real, unit-determinant matrix, hence it belongs to the same symplectic group $\mathrm{Sp}(2,\mathbb{R})$ as $\T$ and the same considerations apply for both matrices (in particular, $\TJ$ has real eigenvalues if $\vert \mathrm{Tr}(\TJ)\vert \geq 2$).

Other bases than the one introduced in \Eq{eq:modes_bessel} can be used. For instance, consider the Hankel basis
\bea
\label{eq:Hankel:basis}
    \tilde{u}^{(1)}_i(k,\tau) 
    & =  \frac{\sqrt{-\pi\tau}}{2}H_{\nu_i}^{(1)}\left(-k\tau\right)e^{i\frac{\pi}{2}\left(\nu_i+\frac{1}{2}\right)} \, ,
    \\
   \tilde{u}^{(2)}_i(k,\tau) 
    &=  \frac{\sqrt{-\pi\tau}}{2}H_{\nu_i}^{(2)}\left(-k\tau\right)e^{-i\frac{\pi}{2}\left(\nu_i+\frac{3}{2}\right)}  \, .
\eea
In this basis, a decomposition analogous to \Eq{eq:expansion:Bessel} can be given,
\bea\label{eq:expansion:Hankel}
    u_{i,k}(\tau)=  \tilde{a}_{i,1}(k) \tilde{u}^{(1)}_i(k,\tau)  + \tilde{a}_{i,2}(k)\tilde{u}^{(2)}_i(k,\tau) \,,
\eea
where the vector coefficient $\tilde{\bm{a}}=(\tilde{a}_{i,1},\tilde{a}_{i,2})$ can be set to the Bunch--Davies vacuum in the first constant-roll phase via
\bea
    \label{eq:atilde:BD}
    \tilde{\bm{a}}_\I(k) = 
    \begin{pmatrix}
    1\\
    0
    \end{pmatrix}\, .
\eea
The vector coefficient is mapped through the transitory phase by a transfer matrix $\TH$, like in \Eq{eq:MS_phases_T},
\be\label{eq:MS_phases_T_H}
     \tilde{\bm{a}}_\II(k)
    = \TH (k)  \tilde{\bm{a}}_\I(k)\, .
\ee
Here, $\TH$ is related to $T$ by a relation analogous to \Eq{eq:TJ}, 
\bea
\label{eq:TH}
    \TH (k) = \WHankel{\II}^{-1}(k)\T (k) \WHankel{\I}(k)\, ,
\eea
where we defined the Wronskian
\bea
\label{eq:W_Hankel}
    \WHankel{i} = W\left(\tilde{u}^{(1)}_i,\tilde{u}^{(2)}_i\right)\, .
\eea
Here, $\tilde{u}^{(1)}_i$ and $\tilde{u}^{(2)}_i$ are the Hankel-basis functions given in \Eq{eq:Hankel:basis}.
They have been normalised such that their Wronskian has unit determinant, hence $\det \TH =1$. However, in that basis the Wronskian matrix is not real, hence $\TH$ does \emph{not} belong to $\mathrm{Sp}(2,\mathbb{R})$.

Nonetheless, one can show that $T_H$ belongs to the group $\mathrm{SU}(1,1)$, which is isomorphic to $\mathrm{Sp}(2,\mathbb{R})$. This can be done as follows. Combining \Eqs{eq:expansion:Bessel} and~\eqref{eq:expansion:Hankel}, the $a$ coefficients in the Bessel and Hankel bases are related via
\bea 
\label{eq:atilde:a}
\tilde{\bm{a}}_i = \underbrace{W^{-1}(\tilde{u}^{(1)}_i,\tilde{u}^{(2)}_i) W(u^{(1)}_i,u^{(2)}_i)}_{{\mathcal{U}}_{\nu_i}}\bm{a}_i\, ,
\eea
which defines the matrix ${\mathcal{U}}_\nu$. Using well-known identities involving the Bessel functions~\cite{NIST:DLMF} one can compute
\bea 
\label{eq:U_nu}
{\mathcal{U}}_\nu= e^{-i\frac{\pi}{4}}
\begin{pmatrix}
\frac{e^{i \frac{\pi}{2}\nu}}{2\sin(\pi\nu)} & -e^{-i \frac{\pi}{2}\nu}\\
-\frac{e^{-i \frac{\pi}{2}\nu}}{2\sin(\pi\nu)} & e^{i \frac{\pi}{2}\nu}
\end{pmatrix}\, .
\eea 
This matrix is the one that relates the transfer matrices in the two bases, via
\bea 
\TH  = {\mathcal{U}}_{\nu_\II} \TJ {\mathcal{U}}_{\nu_\I}^{-1}\, .
\eea 
The identity~\eqref{eq:symplectic:relation} for the $\TJ$ matrix translates into
\bea 
\label{eq:SU11:cond}
 \TH^\dagger {J}_{\nu_\II} \TH   = {J}_{\nu_\I} \,,
\eea 
with ${J}_\nu = {\mathcal{U}}_{\nu}^{-1\dagger} \Omega{\mathcal{U}}_{\nu}^{-1}$. A direct calculation shows that ${J}_\nu$ is independent of $\nu$ and reads
\bea
\label{eq:J:def}
{J}=\begin{pmatrix}
i & 0\\
0 & -i
\end{pmatrix}\, .
\eea 
Together with the condition $\det\TH=1$, \Eq{eq:SU11:cond} defines the group $\mathrm{SU}(1,1)$. It may be viewed as the ``helicity representation''\footnote{The matrix $\mathcal{U}_\nu$ is indeed similar to the matrix $\mathcal{U}$ defining the creation and annihilation operators from the position and momentum operators in quantum mechanics, $\begin{pmatrix} \hat{a}\\ \hat{a}^\dagger \end{pmatrix} = \frac{1}{\sqrt{2}} \begin{pmatrix} 1 & i\\ 1 & -i \end{pmatrix} \begin{pmatrix} \hat{x}\\ \hat{p}   \end{pmatrix} \equiv \mathcal{U}\begin{pmatrix} \hat{x}\\ \hat{p}   \end{pmatrix}  $, with which it even coincides for $\nu=1/2$. They are such that $\mathcal{U}_\nu^{-1 \dagger} \Omega \mathcal{U}_\nu^{-1} = \mathcal{U}^{-1 \dagger} \Omega \mathcal{U}^{-1}=J$. The ladder operators belong to the so-called ``helicity representation'', described by the group $\mathrm{SU}(1,1)$, which explains why transfer matrices in the Hankel basis belong to that group. Another way to stress the correspondence between the Hankel basis and the helicity representation is to notice that, in the Hankel basis, the vacuum state, which is the one annihilated by the annihilation operator, takes the simple form~\eqref{eq:atilde:BD}. The fact that mode functions are real in the Bessel basis is also consistent with the fact that in the position representation, the position and momentum operators are Hermitian, and are mapped via elements of $\mathrm{Sp}(2,\mathbb{R})$.\label{footnote:helicity}}
of $\mathrm{Sp(2,\mathbb{R})}$, to which it is isomorphic~\cite{Grain:2019vnq}, and matrices belonging to $\mathrm{SU}(1,1)$ are of the form
\bea 
\TH = \begin{pmatrix}
\alpha & \beta\\
\beta^* & \alpha^*
\end{pmatrix}, 
\qquad \mbox{with} \qquad
\vert\alpha\vert^2-\vert\beta\vert^2=1.
\eea 
The parameters $\alpha$ and $\beta$ are often called ``Bogoliubov coefficients''.

\subsection{Power spectrum}

As mentioned above, the power spectrum of the curvature perturbations at late times,
\be
    \mathcal{P}_\zeta(k) = \frac{k^3}{2\pi^2}\left|\frac{u_k}{z}\right|^2_{\tau \to 0}\, ,
\ee 
is determined by the growing mode in the last constant-$\nu$ phase. Expanding the Bessel functions at leading order in phase $\II$, we find
\bea
\label{eq:calP:interm}
    \mathcal{P}_\zeta(k) 
    = \frac{\Gamma^2(\nu_\II )}{ 2\pi^3} 2^{2\nu_\II-2} k^2  \left|\frac{a_{\II,1}(k)}{z}\right|^2 \left(-k \tau\right)^{1-2\nu_\II} \bigg|_{\tau\to 0}\, .
\eea
It is clear that $z$ satisfies the Mukhanov--Sasaki equation for $k=0$ and in the regime $\epsilon_1\ll 1$ where $a\simeq -1/(H\tau)$, its solution is of the form $z=z_{\II,1} a^{\nu_\II - 1/2} + z_{\II,2} a^{-\nu_\II - 1/2}$. This also corresponds to taking the limit $k\to 0$ in \Eq{eq:expansion:Bessel}. At late time the growing mode dominates, $z=z_{\II,1} (-H_\II \tau)^{1/2-\nu_\II}$, where $H_\II$ is the (approximately constant) value of the Hubble parameter during the late constant-roll phase. Inserting this expression into \Eq{eq:calP:interm} shows that the power spectrum asymptotes to a constant at late times,
\bea
\label{eq:Pzeta:gen}
    \mathcal{P}_\zeta(k)
    = \frac{\Gamma^2(\nu_\II )}{ 4 \pi^3} k^2  \left|\frac{a_{\II,1}(k)}{z_{\II,1}}\right|^2 \left(\frac{k}{2H_\II}\right)^{1-2\nu_\II}\, .
\eea
Using \Eq{eq:atilde:a}, the power spectrum can also be written in terms of the $\tilde{a}$ coefficients in the Hankel basis, since
\bea
\label{eq:a_II_in_Hankel}
    \vert a_{\II,1}(k)\vert^2 = \vert \tilde{a}_{\II,1}(k)+e^{-i\pi\nu_\II} \tilde{a}_{\II,2}(k)\vert^2 \, .
\eea
Both $\tilde{a}_{\II,1}(k)$ and $\tilde{a}_{\II,2}(k)$ appear in the power spectrum in the Hankel basis, since the growing mode receives contributions from both (the Hankel decomposition is not onto growing and decaying modes).

Note that, in the scenario described around \Fig{fig:schematic plot}, modes exiting early in phase I and in the $\II_{\mr{CR}}$ phase are frozen after Hubble exit and can thus be evaluated via matching at $k=\mathcal{H}$, where $\mathcal{H}=a'/a=aH$. In general, we can show that this yields the following asymptotes:
\be\label{eq:Pzeta_asymp}
    \mathcal{P}_\zeta(k) \sim 
\left\{
\begin{array}{ll}    
    \mathcal{P}^{(\mr{CR})}_\zeta(k;\nu_\I) & \quad \mbox{when} \quad k \ll \mathcal{H}_{\I}  \\
    \mathcal{P}^{(\mr{CR})}_\zeta(k;\nu_\II) \times |a_{\II,1}(k)|^2  & \quad \mbox{when} \quad k \gg \mathcal{H}_{\II} \\
\end{array}
\right.\, ,
\ee
where 
\be
\label{eq:CR Pzeta}
    \mathcal{P}^{\rm (CR)}_\zeta(k;\nu) = \left. \frac{\Gamma(\nu)^2}{\pi^3 4^{2-\nu}}  \frac{H^2}{\Mp^2 \eps_1}   \right|_{k=\mathcal{H}} \propto k^{3-2\nu_{\II}}
\ee
gives the spectrum obtained in a purely constant-roll phase~\cite{Martin:2012pe, Motohashi:2014ppa, Motohashi:2019rhu, Inui:2024sce}, starting with a Bunch--Davies vacuum. This confirms that modes exiting early in phase $\I$ follow the usual constant-roll spectrum, as expected from scale decoupling. Modes exiting close to or after the beginning of the $\II_{\mr{CR}}$ phase will be found to be modulated by $|a_{\II,1}(k)|^2$, which is an imprint of non-trivial sub-Hubble (or near-Hubble) evolution. We will show later that, for sufficiently smooth transitions, $|a_{\II,1}(k)|^2 \to 1$ when $k \gg \mathcal{H}_{\II}$, that is, scale decoupling is also satisfied for large-$k$ modes.

In what follows, we thus study the two ranges of scales where the power spectrum is not merely given by the constant-roll formula~\eqref{eq:CR Pzeta}: the modes that exit the Hubble radius at the beginning of phase $\II$ (\Sec{sec:subhorizon}) and the modes that exit at the end of phase $\I$ (\Sec{sec:suphorizon}), see the right panel of \Fig{fig:schematic plot}. In practice, we will compute the matrices $T_J$ and $T_H$ in these limits, from which we will extract the coefficients $a_{\II,i}(k)$ and $\tilde{a}_{\II,i}(k)$ and thus the power spectrum using \Eqs{eq:MS_phases_T} and \eqref{eq:MS_phases_T_H}. With the Bunch--Davies initial conditions \eqref{eq:a:BD} and \eqref{eq:atilde:BD}, we have
\bea
\label{eq:a_II_in_T}
    \left|a_{\II,1}(k)\right|^2 &=\left|  \TJ{}_{11}(k) + \frac{e^{-i\pi \nu_\I}}{2\sin(\pi\nu_\I)} \TJ{}_{12}(k)\right|^2 \\
    &= \left| \TH{}_{11}(k) + e^{-i\pi \nu_\II} \TH{}_{21}(k)\right|^2 \, .
\eea
The assumption of a Bunch--Davies initial state may not hold in scenarios where additional features are generated before phase $\I$. We will, however, not consider such scenarios here.

\section{Damping of oscillations at small scales}
\label{sec:subhorizon}

In this section, we consider the scales that exit the Hubble radius at the beginning of phase $\II$. They lie to the right of the power-spectrum peak in \Fig{fig:schematic plot}. These scales are only mildly sub-Hubble during the transition, hence they are subject to non-trivial sub-Hubble evolution that gives rise to the damped oscillations in the power spectrum which we want to characterise.

In the sub-Hubble regime, the $k^2$ term dominates in the Mukhanov--Sasaki equation~\eqref{eq:MS}, and the $z''/z$ term can be treated perturbatively. The leading-order (flat space-time) solutions correspond to plane waves, and non-trivial features, such as the power-spectrum oscillations, appear at next-to-leading order. This is why we start by computing the transfer matrix~\eqref{eq:MS_phases_T} at next-to-leading order in $z''/z$.

\subsection{Perturbative transfer matrix}
\label{sec:subH:exp}

At leading order in $z''/z$, the two independent solutions of the Mukhanov--Sasaki equation \eqref{eq:MS} read\footnote{The Wronskian determinant for these modes is $\det W = -2 i k$. This makes the normalization of $G_{\mathrm{sub}}$ in \Eq{eq:MS_integral_large_k} differ slightly from the conventional one and introduces explicit powers of $1/k$ in the expansion \eqref{eq:mode expansion large k}.}
\be\label{eq:u_0 large k}
    u^{(0)}_{1,\mathrm{sub}}(\tau) = e^{i k\tau}, \qquad
    u^{(0)}_{2,\mathrm{sub}}(\tau) = e^{-i k\tau}\, .
\ee
The index ``$\mathrm{sub}$'' refers to sub-Hubble, as these solutions are valid only when $k \gg \mathcal{H}_\II$. The Mukhanov--Sasaki equation \eqref{eq:MS} can be rewritten as an integral equation
\bea\label{eq:MS_integral_large_k}
    u_k(\tau) 
    & = u^{(0)}_{\mathrm{sub}}(\tau) - \frac{1}{2ki} \int^{\tau}_{\tau_{\I}} \td \tau'  G_{\mathrm{sub}}(\tau,\tau') u_k(\tau')\frac{z''}{z}(\tau'), 
    \qquad 
    u^{(0)}_{\mathrm{sub}} \equiv \sum_{i = 1,2} b_{i} u^{(0)}_{i,\mathrm{sub}}(\tau)\, ,
\eea
where the coefficients $b_i$ are determined from the initial conditions set at $\tau=\tau_\I$, and the (retarded) Green function at small scales reads
\bea
    G_{\mathrm{sub}}(\tau,\tau') 
    &= \theta(\tau-\tau') \left[u^{(0)}_{2,\mathrm{sub}}(\tau) u^{(0)}_{1,\mathrm{sub}}(\tau') - u^{(0)}_{1,\mathrm{sub}}(\tau) u^{(0)}_{2,\mathrm{sub}}(\tau')\right]\, \\
    &= \theta(\tau-\tau') 2i \sin\left[k(\tau'-\tau)\right]\, .
\eea
Iterating the integral equation~\eqref{eq:MS_integral_large_k} leads to a series expansion in $z''/z$,
\bea\label{eq:mode expansion large k}
    u_k(\tau) 
    = & u^{(0)}_{\mathrm{sub}} - \frac{1}{2ki} \int^{\tau}_{\tau_{\I}} \td \tau'  G_{\mathrm{sub}}(\tau,\tau') u^{(0)}_{\mathrm{sub}}(\tau') \frac{z''}{z}(\tau')
    + \ldots = \sum_{n\geq 0} (-2ki)^{-n} u^{(n)}_{\mathrm{sub}}(\tau),
    \\ & 
    \quad \mbox{with} \quad
    u^{(n)}_{\mathrm{sub}}(\tau_0) \equiv \left[\prod^{n}_{j=1} \int^{\tau_{j-1}}_{\tau_{\I}} \td \tau_j G_{\mathrm{sub}}(\tau_{j-1},\tau_{j})\frac{z''}{z}(\tau_j)\right] u^{(0)}_{\mathrm{sub}}(\tau_n)\, .
\eea
In particular, two independent solutions $u_{1,\mathrm{sub}},u_{2,\mathrm{sub}}$ can be constructed at any order in $z''/z$ by choosing $u^{(0)}_{\mathrm{sub}} = u^{(0)}_{1,\mathrm{sub}}$ (that is, $b_1=1,b_2=0$), or $u^{(0)}_{\mathrm{sub}} = u^{(0)}_{2,\mathrm{sub}}$ ($b_1=0,b_2=1$).

These solutions give rise to an expansion of the transfer matrix via the formula $\Wsub(\tau_\II)=T \Wsub(\tau_\I)$, see \Eq{eq:T:in:W:and:W-1}. Here, $\Wsub$ denotes the Wronskian matrix in the basis $(u_{1,\mathrm{sub}},u_{2,\mathrm{sub}})$, and $T$ is the transfer matrix in that same basis. We get
\bea\label{eq:T_in_W}
T=\Wsub(\tau_\II) W^{-1}_{\mathrm{sub}}(\tau_\I) \, .
\eea 
This can also be cast into an expression of the transfer matrix in the Hankel basis, which proves convenient at small scales since the two independent solutions $\tilde{u}^{(1)},\tilde{u}^{(2)}$ defined in \Eq{eq:Hankel:basis} asymptote to the plane-wave solutions~\eqref{eq:u_0 large k}. Using \Eq{eq:TH}, one finds
\bea\label{eq:TH_in_W}
T_H(k) = \WHankel{\II}^{-1}(k) \Wsub(\tau_\II) W^{-1}_{\mathrm{sub}}(\tau_\I) \WHankel{\I}(k)\, .
\eea 
By construction, $\Wsub(\tau_\I) = W(u_{1,\mathrm{sub}}^{(0)},u_{2,\mathrm{sub}}^{(0)})$,
where $u_{i,\mathrm{sub}}^{(0)}$ are given in \Eq{eq:u_0 large k}, and is thus straightforward to compute. One can then use \Eq{eq:mode expansion large k} to expand $\Wsub(\tau_\II)$. For $\WHankel{\I}(k)$ and $\WHankel{\II}(k)$, using \Eq{eq:mode expansion large k} amounts to expanding the Hankel functions appearing in \Eqs{eq:Hankel:basis} in powers of $k \tau$. At next-to-leading order,\footnote{Although the expansions of $\Wsub(\tau_\II)$ and of $\WHankel{i}(k)$ are both controlled by $z''/z$, either during the transition or at its end points, they are technically distinct and in practice they are carried out independently. The expansion of $\Wsub(\tau_\II)$ gives rise to the $\tilde{A}$ terms in \Eq{eq:T_sub_pert}, while $\WHankel{i}(k)$ gives rise to the $\tilde{B}$ terms. We set the order of these expansions such that the first non-vanishing contribution in all entries of $\delta T_H$ is obtained.} this gives rise to 
\bea\label{eq:T_sub_pert}
    \TH(k) 
    = &  1 + \delta \TH + \mathcal{O}(z''/z)^2 \\
    = & 1 + 
\frac{1}{2ik}
\begin{pmatrix}
-\tilde{A}(0) + \tilde{B}_\I(0)- \tilde{B}_\II(0) & \quad & i \tilde{A}(2k)+\frac{\tilde{B}_\I(2k)}{2k\tau_\I}-\frac{\tilde{B}_\II(2k)}{2k\tau_\II}\\
i \tilde{A}(-2k)+\frac{\tilde{B}_\II(-2k)}{2k\tau_\II}-\frac{\tilde{B}_\I(-2k)}{2k\tau_\I} & & \tilde{A}(0)+\tilde{B}_\II(0)-\tilde{B}_\I(0)
\end{pmatrix}  + \mathcal{O}(z''/z)^2\, ,
\eea
where
\be
\label{eq:A(k):def}
    \tilde A(k) \equiv \int^{\tau_\II}_{\tau_\I} \td \tau' e^{i k\tau'}\frac{z''}{z}(\tau') 
    \quad\text{and}\quad 
    \tilde{B}_i(k)=\frac{e^{ik\tau_i}}{\tau_i} \left(\nu_i^2-\frac{1}{4}\right)
    \,.
\ee
Since $\tilde{A}(-k)=\tilde{A}^*(k)$ and $\tilde{B}(-k)=\tilde{B}^*(k)$, one can show that $(\delta T_H)^\dagger J + J \delta T_H=0$, where $J$ was defined in \Eq{eq:J:def}. This implies that $\delta T$ belongs to the Lie algebra of $\mathrm{SU(1,1)}$, in agreement with the discussion in \Sec{sec:connect:two:constant:nu}.

The elements of the $\delta T_H$ matrix are manifestly invariant when shifting $\tau_\I$ and $\tau_\II$ in the constant-$\nu_\I$ and constant-$\nu_\II$ phases respectively. This is because, within these phases, $z''/z=(\nu_i^2-1/4)/\tau^2$, and using this in \Eqs{eq:A(k):def} and \eqref{eq:T_sub_pert} we get $\partial T_H/\partial \tau_i=0$ at the order at which the calculation is performed. 
To put this in another way, if $\tau_\II$ and $\tau_\II'$ both belong to the $\II$ constant-$\nu$ phase, $T_{H,\tau_\I\to\tau_\II} =T_{H,\tau_\II'\to\tau_\II}  T_{H,\tau_\I\to\tau_\II'}$ where $T_{H,\tau_\II'\to\tau_\II}=1$ since it maps the evolution within a constant-$\nu$ phase. This implies that $T_{H,\tau_\I\to\tau_\II}$ is independent of $\tau_\II$ as long as $\tau_\II$ belongs to the constant-$\nu_\II$ phase, and likewise it does not depend on  $\tau_\I$ as long as it belongs to the constant-$\nu_\I$ phase.

\subsection{Power spectrum}
\label{sec:sub_hubble_power_spectrum}

Combining \Eqs{eq:Pzeta_asymp}, \eqref{eq:a_II_in_T} and~\eqref{eq:T_sub_pert}, the power spectrum at next-to-leading order in $z''/z$ is given by
\bea
\label{eq:Pzeta large k first correction}
   \frac{\mathcal{P}_\zeta(k)}{ \mathcal{P}^{\mathrm{ (CR)}}_\zeta(k;\nu_\II)}&=
   1+\frac{1}{k}\Re \left\{ e^{i\pi\nu_\II} \left[ \tilde{A}(2k)+i \frac{\tilde{B}_\II(2k)}{2k\tau_\II}-i \frac{\tilde{B}_\I(2k)}{2k\tau_\I}\right]\right\}  
     \, .
\eea
Notice that diagonal terms in \Eq{eq:T_sub_pert} do not bring any contribution to the power spectrum at first order in the perturbative scheme described above.

In the second term on the right hand side of \Eq{eq:Pzeta large k first correction}, the function $\tilde{A}(k)$ is rotating in the complex plane, leading to oscillations in the power spectrum. The second term becomes small for large $k$: the oscillations get damped and the power spectrum asymptotes to the CR result. We now turn to the description of the damping of the oscillations.

The $\tilde{A}$ integral can be further expanded in inverse powers of $k$ by performing partial integration in \Eq{eq:A(k):def}, leading to
\be
\label{eq:A(k):IntByPart}
    \tilde A(k) = \underbrace{\frac{e^{i k\tau}}{i k}\frac{z''}{z}}_{\tilde{B}(k)/(ik\tau)}\Bigg|^{\tau_\II}_{\tau_\I}  - \frac{1}{i k}\int^{\tau_\II}_{\tau_\I} \td \tau' e^{i k\tau'}\frac{\td}{\td \tau}\left[\frac{z''}{z}(\tau')\right] .
\ee
Here, one notices that the $\tilde{B}$ terms in \Eq{eq:Pzeta large k first correction} exactly cancel out the boundary contributions of the $\tilde{A}$ integrals. This implies that, although $\tilde{A}(k)/k$ is a priori of order $k^{-1}$, when $z''/z$ is differentiable the correction to the power spectrum is suppressed by $k^{-2}$. 

The same procedure can be iterated: through repeated integration by parts, when $z$ is smooth, all power-law suppression of $\tilde A(k)$ can be transferred to boundary terms containing derivatives of $z''/z$. In practice, the $\tilde B(k)$ contributions are obtained by expanding the Hankel functions appearing in the $\tilde{W}_i(k)$ in powers of $k\tau$. In \Eq{eq:A(k):def} that was done to next-to-leading order, but if the expansion was performed to higher order one would cancel out all boundary terms stemming from integrating $\tilde A(k)$ by part, order by order. This is in fact necessary for the $T_H$ matrix to be independent of $\tau_\I$ and $\tau_\II$, see the discussion at the end of \Sec{sec:subH:exp}.

As a consequence, if $z$ is smooth, then the correction to the power spectrum falls off more rapidly than any inverse power of $k$, \ie it decays non-perturbatively.
This is a consequence of the Riemann--Lebesgue lemma, which states that the Fourier transform $f_k$ of an integrable function vanishes as $k\to\infty$. If the function can be differentiated $n$ times (so that all the derivatives are integrable), the Fourier transforms of the derivatives must similarly vanish; the highest of them is $(ik)^n f_k$, so $f_k$ must vanish faster than $k^{-n}$. In our case,  $\tilde{A}(k)$ may be seen as the Fourier transform of $z''/z$, from which $\tilde{B}$ subtracts the boundary contributions. If $z''/z$ can be differentiated $n$ times but not $n+1$ times, then the correction to the power spectrum is of order $k^{-(n+1)}$. This gives rise to two different behaviours, depending on how $k$ compares to $1/\Delta\tau$, where $\Delta\tau$ is the characteristic duration of the pulse in $z''/z$:

\begin{itemize}
    
\item  When $k \Delta \tau \gg 1$, $z''/z$ and all its derivatives appear smooth; the power spectrum oscillations decay faster than any inverse power of $k$. We expect this damping to be exponential: $\mathcal{P}_\zeta/\mathcal{P}_\zeta^{\mathrm{CR}} -1 \propto \Re (e^{-\alpha k \Delta \tau})$, where $\alpha$ is some dimensionless coefficient. This behaviour will be found in the example studied in \Sec{sec:example_damping}. 

\item  When the pulse is sharp compared to $1/k$, \ie $k\Delta\tau\ll 1$, the derivatives of $z''/z$ do not appear smooth. \Eq{eq:A(k):def} gives $\tilde A(2k) \approx \tilde A(0) e^{2ik\tau_0}$, with $\tau_0$ denoting the time of the transition. The correction to the power spectrum displays a slow power-law damping, with oscillations at the frequency $2\tau_0$. This leads to the  $k^{-1}$ behaviour found in the limit studied in \Sec{sec:instantaneous_transitions}.

\end{itemize}

\begin{figure}[t]
\centering
\includegraphics[height=0.4\textwidth,trim={0cm -0.8cm 0cm 0cm}, clip]{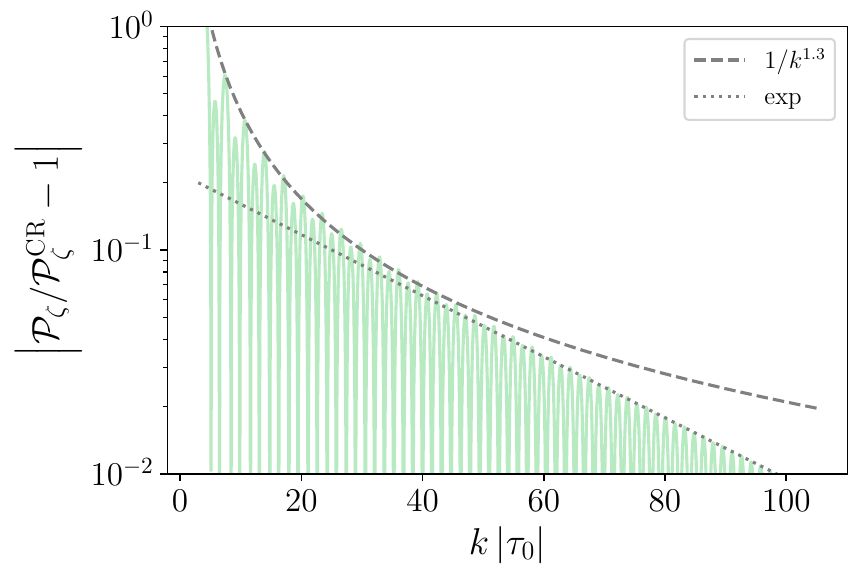}
\caption{Damped oscillations of the power spectrum at large $k$ in the model studied in \Sec{sec:smooth_tanh}, see \Eq{eq:smooth tanh} with parameters $\nu_{\I}=3/2$, $\nu_{\II}=2$, $A = 3.54$ and $\tau_0/\Delta\tau = 100$. When $k\left|\tau_{0}\right|\le 40$, oscillations are fitted by a power law $\sim k^{-1.3}$, resulting from the presence of two damped contributions with powers $1/k$ and $1/k^2$ respectively. At larger values of $k\left|\tau_{0}\right|$, oscillations are exponentially damped.}
\label{fig:transition damping}
\end{figure}

Notice that in the case of a sharp, though not instantaneous transition, both the power-law regime at $\tau_\II^{-1}\ll k\ll \Delta \tau^{-1}$ and the exponential regime at $k\gg \Delta \tau^{-1}$ may be found, see the example displayed in \Fig{fig:transition damping}. 

\section{Existence of a dip and its relation to the peak}
\label{sec:suphorizon}

Instead of treating $z''/z$ as a small perturbation in the Mukhanov--Sasaki equation as was done in the previous section, we can consider the opposite limit where the solutions are expanded in powers of $k^2$. This allows us to treat modes that exit the Hubble radius near the end of phase $\I$, since they are (mildly) super-Hubble during the transition. In \Fig{fig:schematic plot}, these are the modes to the left of the power spectrum peak, exhibiting a possible downward dip followed by steep growth. Through our perturbative analysis, we will find a criterion for the existence of the dip and derive a relation between its amplitude and the amplitude of the peak that occurs at shorter scales.

\subsection{Expansion of the mode functions}

For super-Hubble modes, the $k^2$ term in the Mukhanov--Sasaki equation \eqref{eq:MS} can be treated perturbatively~\cite{Leach:2001zf}. At leading order, one obtains the so-called growing- and decaying-mode solutions\footnote{In the definition of $u^{(0)}_{2,\mathrm{sup}}(\tau)$, the integration constant has been set so that the decaying mode vanishes at late time, although other choices are possible~\cite{Jackson:2023obv}.}
\be\label{eq:u_0}
    u^{(0)}_{1,\mathrm{sup}}(\tau) = z(\tau), \qquad
    u^{(0)}_{2,\mathrm{sup}}(\tau) = z(\tau) \, \underline{\mathcal{H}}\!\!\int^{\tau}_{0} \frac{\td \tau'}{z^2(\tau')}\, ,
\ee
where $\underline{\mathcal{H}}$ stands for Hadamard regularisation which must be used to eliminate singularities in the integral if $z$ crosses 0 (see Appendix~\ref{app:Hadamard} for details). The first mode leads to a frozen curvature perturbation $\zeta=u/z$, while the time behaviour of the second mode is model dependent. The Wronskian determinant for these solutions is normalised to $\det W = u^{(0)}_{1,\mathrm{sup}} u^{(0)\prime}_{2,\mathrm{sup}}-u^{(0)}_{2,\mathrm{sup}}u^{(0)\prime}_{1,\mathrm{sup}} = 1$.

Analogously to \Eq{eq:MS_integral_large_k}, the Mukhanov--Sasaki equation~\eqref{eq:MS} can be recast as an integral equation
\bea\label{eq:MS_integral_k}
    u_k(\tau) 
    & = u^{(0)}_{\mathrm{sup}}(\tau) - k^2 \int^{\infty}_{\tau_{\I}} \td \tau'  G_{\mathrm{sup}}(\tau,\tau') u_k(\tau'), 
    \qquad 
    u^{(0)}_{\mathrm{sup}} \equiv \sum_{i = 1,2} c_{i} u^{(0)}_{i,\mathrm{sup}}(\tau)\, ,
\eea
where $c_i$ are determined from the initial conditions at $\tau_{\I}$ and the Green function is given by
\bea
    G_{\mathrm{sup}}(\tau,\tau') 
    &= \theta(\tau-\tau') \left[u^{(0)}_{2,\mathrm{sup}}(\tau) u^{(0)}_{1,\mathrm{sup}}(\tau') - u^{(0)}_{1,\mathrm{sup}}(\tau) u^{(0)}_{2,\mathrm{sup}}(\tau')\right]\, \\
    &= \theta(\tau-\tau') z(\tau)z(\tau') \int^{\tau}_{\tau'} \frac{\td \tau''}{z^2(\tau'')}.
\eea
The solution can again be formally expressed by the series expansion
\bea\label{eq:MS_exp_small_k}
    u_k(\tau) 
    &= u^{(0)}_{\mathrm{sup}}(\tau) - k^2 \int^{\tau}_{\tau_{\I}} \td \tau'  G_{\mathrm{sup}}(\tau,\tau') u^{(0)}_{\mathrm{sup}}(\tau') 
    + \ldots \\
    &= \sum_{n\geq 0} (-k^{2})^n u^{(n)}_{\mathrm{sup}}(\tau), 
    \quad \mbox{with} \quad
    u^{(n)}_{\mathrm{sup}}(\tau_0) \equiv \left[\prod^{n}_{j=1} \int^{\tau_{j-1}}_{\tau_{\I}} \td \tau_j G_{\mathrm{sup}}(\tau_{j-1},\tau_{j})\right] u^{(0)}_{\mathrm{sup}}(\tau_n)\, .
\eea
Two linearly-independent solutions $u_{i,\mathrm{sup}}$ can be constructed perturbatively at any order in $k^2$ by choosing $u^{(0)}_{\mathrm{sup}} = u^{(0)}_{i,\mathrm{sup}}$. 

\sloppy Let us note that the Wronskians of $u_{i,\mathrm{sup}}(\tau)$ and $u^{(0)}_{i,\mathrm{sup}}$ have the same determinant, $\mathrm{det}[W(u_{1,\mathrm{sup}},u_{2,\mathrm{sup}})] = \mathrm{det} 
[W(u^{(0)}_{1,\mathrm{sup}},u^{(0)}_{2,\mathrm{sup}})]$. This is because they are constant and, by \Eq{eq:MS_integral_k}, they are equal at $\tau = \tau_\I$, since $u_{i,\mathrm{sup}}(\tau_{\I}) = u^{(0)}_{i,\mathrm{sup}}(\tau_{\I})$ and  $u'_{i,\mathrm{sup}}(\tau_{\I}) = (u^{(0)}_{i,\mathrm{sup}})'(\tau_{\I})$. This implies that the Wronskian determinant and the determinant of the transfer matrix $T$
introduced in \Eq{eq:MS_phases_tildeT} do not receive corrections at any order in $k^2$.\footnote{The Wronskian determinant of the full solutions $u_{i,\mathrm{sup}}(\tau)$ obeys
\bea\label{}
    \frac{\mathrm{det}[W(u_{1,\mathrm{sup}},u_{2,\mathrm{sup}})]}{\mathrm{det} 
[W(u^{(0)}_{1,\mathrm{sup}},u^{(0)}_{2,\mathrm{sup}})]} 
    &= 1 - k^2 (I_{12} - I_{21}) + k^4 (I_{11}I_{22}-I_{12}I_{21}) \\
    &= 1 + \sum_{n \geq 0}(-k^{2})^{n+2}
    \left[ I^{(n+1)}_{12} - I^{(n+1)}_{21} + \sum^{n}_{m=0} \left(I^{(n-m)}_{11}I^{(m)}_{22}-I^{(n-m)}_{12}I^{(m)}_{21}\right) \right] \, ,
\eea
where $I_{ij}(\tau) \equiv \int^{\tau}_{\tau_{\I}} \td \tau'  u^{(0)}_{i,\mathrm{sup}}(\tau') u_{j,\mathrm{sup}}(\tau')$ and $I^{(n)}_{ij}(\tau) \equiv \int^{\tau}_{\tau_{\I}} \td \tau'  u^{(0)}_{i,\mathrm{sup}}(\tau') u^{(n)}_{j,\mathrm{sup}}(\tau')$. As the two Wronskian determinants are equal, each term of the series must vanish, implying that
\be
    I^{(n+1)}_{12} - I^{(n+1)}_{21} + \sum^{n}_{m=0} \left(I^{(n-m)}_{11}I^{(m)}_{22}-I^{(n-m)}_{12}I^{(m)}_{21}\right) = 0\,.
\ee
This relation, analogous to an ``optical theorem'' for the integrals $I_{ij}$, may be used to speed up their computation if needed.
}

Let us then turn to the transfer matrix.
This time, we choose to present the matrix in the Bessel basis, as it features one growing mode that dominates at late time, making the analysis of the power spectrum simple. We stress that we could also work in the Hankel basis, and the final result for the power spectrum would be the same. 
The two linearly-independent solutions $u_{i,\mathrm{sup}}$ give rise to an expansion of the transfer matrix via the formula $\Wsup(\tau_\II)=T \Wsup(\tau_\I)$, see  \Eq{eq:T:in:W:and:W-1}. Here, $\Wsup$ denotes the Wronskian matrix in the basis $(u_{1,\mathrm{sup}},u_{2,\mathrm{sup}})$, and $T$ is the transfer matrix in that same basis. We get
\bea\label{eq:T_in_W_sup}
T=\Wsup(\tau_\II) W^{-1}_{\mathrm{sup}}(\tau_\I) \, .
\eea 
Using \Eq{eq:TJ}, the Bessel-basis transfer matrix is given by
\bea\label{eq:TJ_in_W}
\TJ(k) = \WBessel{\II}^{-1}(k) \Wsup(\tau_\II) W^{-1}_{\mathrm{sup}}(\tau_\I) \WBessel{\I}(k)\, ,
\eea 
which is the equivalent in the Bessel basis of \Eq{eq:TH_in_W}. We can then expand the mode functions and the matrices $\WBessel{i}$ in \Eq{eq:T_in_W_sup} to obtain the expansion of the transfer matrix at large scales.

\subsection{Power spectrum at large scales}
\label{sec:existence_of_dip}

At very large scales, the power spectrum is well described by its CR formula \eqref{eq:CR Pzeta}. However, the possible dip and the peak of the power spectrum do not belong to this regime. This is expected, as both the dip and the peak are consequences of the existence of a non-CR phase -- for instance, a transition into USR. To describe this part of the spectrum, one needs to include the next-to-leading order terms in the $k^2$ expansion described above.

Equations \eqref{eq:Pzeta:gen} and \eqref{eq:a_II_in_T} give the power spectrum in terms of the transfer matrix $T_J$ assuming an initial Bunch--Davies vacuum. Expanding the transfer matrix~\eqref{eq:TJ_in_W} in powers of $k^2$, the power spectrum can be expressed as\footnote{The extra power of $2\nu_I$ in the last term comes from mixing the two mode functions in the non-diagonal term in \Eq{eq:a_II_in_T}.}
\be
\label{eq:Pzeta expansion small k}
   \mathcal{P}_\zeta(k) 
    = \mathcal{P}^{\rm (CR)}_\zeta(k;\nu_\I)\left|1 + \sum_{n \geq 1} A_{2n}\left(\frac{k}{\kone}\right)^{2n}+e^{-i\pi \nu_\I}\sum_{m\geq0} B_{2m}\left(\frac{k}{\kone}\right)^{2m+2\nu_\I}\right|^{2} \, ,
\ee
where $\kone=-1/\tau_\I$ is a reference scale introduced to make the expansion coefficients $A_i$ and $B_i$ dimensionless. These coefficients can be computed using \Eqs{eq:MS_exp_small_k} and \eqref{eq:modes_bessel} and explicit expressions are given below.
Notice that all the coefficients $A_i,B_i$ are real since the transfer matrix is real in the Bessel basis, hence the only imaginary contribution to the quantity inside the modulus may come from the coefficient $e^{-i\pi \nu_\I}$. Since it is further suppressed by $k^{2\nu_\I}$, the real part dominates over the imaginary part when $k\ll k_1$. This is why, although the power spectrum never vanishes exactly as discussed in \Sec{sec:Modulus:MSvariable}, the vanishing of the real part signals the presence of a dip. For this reason, we let $\kdip$ denote the solution (if it exists) of
\be
\label{eq:condition dip}
    1 + \sum_{n\geq1} \left[A_{2n}\left(\frac{\kdip}{\kone}\right)^{2n}+\cos(\pi \nu_\I) B_{2n}\left(\frac{\kdip}{\kone}\right)^{2n+2\nu_\I}\right] = 0\, .
\ee
As we will check below, $\kdip$ provides a reliable approximation for the location of the dip when it exists, while the imaginary part at $\kdip$ properly approximates the amplitude of the power spectrum at its minimum,
\be
\label{eq:amplitude dip}
   \mathcal{P}_\zeta(\kdip) = \mathcal{P}^{\rm (CR)}_\zeta(\kdip;\nu_\I)\sin^2(\pi \nu_\I)\left(\frac{\kdip}{\kone}\right)^{4\nu_\I}\left[\sum_{m\geq0} B_{2m}\left(\frac{\kdip}{\kone}\right)^{2m}\right]^{2}\,.
\ee
Computing the first coefficients $A_i$, $B_i$ gives
\be
\label{eq:Different coefficients expansion small k}
\begin{split}
    \frac{A_2}{\kone^2} 
    =& \frac{\tau_\I^2}{4\left(\nu_\I-1\right)} - \underline{\mathcal{H}}\left[\int_{\tau_\I}^0 \frac{\td \tau'}{z^2(\tau')} \int_{-\infty}^{\tau'}\td \tau''z^2(\tau'')\right] \, , 
    \\
    \frac{A_4}{\kone^4} 
    =& \frac{1}{8\left(\nu_\I-1\right)^2}\left[ - \frac{\tau_\I^4}{4\left(2-\nu_\I\right)} - \tau_\I^2u^{(0)}_{1,\mathrm{sup}}(\tau_\I)^2u^{(0)}_{2,\mathrm{sup}}(\tau_\I)^2 + \frac{1}{\left(2-\nu_\I\right)}\tau_\I^3u^{(0)}_{1,\mathrm{sup}}(\tau_\I)u^{(0)}_{2,\mathrm{sup}}(\tau_\I)\right] 
    \\
    &- \int_{\tau_\I}^0\td \tau'u^{(0)}_{2,\mathrm{sup}}(\tau')^2\int_{-\infty}^{\tau'}\td \tau''u^{(0)}_{1,\mathrm{sup}}(\tau'')^2 + \frac{1}{2\kone^4}A_2^2 \, , 
    \\
    \frac{B_0}{\kone^{2\nu_\I}} 
    =& \left(\frac{-\tau_\I}{2}\right)^{2\nu_\I}\frac{\Gamma\left(1-\nu_\I\right)}{\Gamma\left(1+\nu_\I\right)}\left[\frac{2\nu_\I u^{(0)}_{1,\mathrm{sup}}(\tau_\I)u^{(0)}_{2,\mathrm{sup}}(\tau_\I)}{\tau_\I}-1\right] \, .
\end{split}
\ee
Despite appearances, these formulae are independent of $\tau_\I$ as long as it belongs to the initial CR phase. Hereafter, it is convenient to take $\tau_\II$ to zero, \ie at the end of inflation.

Even though the integrals above are well defined (assuming one computes the Hadamard part if needed), they are not easy to evaluate numerically. In Appendix~\ref{app:Numerical formulae}, we show how those coefficients can be written in terms of integrals between $\tau_\I$ and $\tau_\II$ that can be evaluated without any numerical issue. There, we take $\tau_\II$ to be right after the feature in $z''/z$ instead of at the end of inflation, in order to efficiently capture contributions from the transition region around the feature. 

\subsection{Criterion for the existence of a dip}
\label{sec:dip_crit}

In the following, we assume that the power spectrum features a pronounced peak, so that large fluctuations, perhaps leading to primordial black holes, can form. This means that $\abs{z}$, \ie the inflaton velocity, drops significantly during the USR phase, see \Fig{fig:schematic plot}, and remains small for sufficiently long in order to enhance the curvature fluctuations, before starting to grow again in the second CR phase. As $z$ decreases it may even cross zero and flip sign, which has important consequences for the power spectrum as we will see below. We further assume that $\nu_{\I}=3/2$, so that phase $\I$ proceeds in the slow-roll regime for simplicity, but the results derived below can be easily extended to a generic $\nu_{\I}$. Notice that under this assumption, $\mathcal{P}^{\rm (CR)}_\zeta(k;\nu_{\I})$ becomes independent of $k$ (a perfectly scale-invariant spectrum) and we call it $\mathcal{P}_\zeta^{\mathrm{CMB}}$ since the scales at which the CMB is measured are expected to fall in the regime where the power spectrum is well-approximated by $\mathcal{P}^{\rm (CR)}_\zeta(k;\nu_{\I})$. For numerical results, we take $\mathcal{P}_\zeta^{\mathrm{CMB}}= 2.1\times10^{-9}$ in agreement with CMB measurements~\cite{Akrami:2018odb, BICEP:2021xfz}. 

Below \Eq{eq:Different coefficients expansion small k}, we stressed that the coefficients obtained in the large-scale expansion are independent of $\tau_\I$, but here we explicitly choose $\tau_\I$ to be the time at which the feature starts, that is, the conformal time at the end of the initial SR phase. Since we consider cases where the power spectrum is strongly enhanced, this allows us to assume that the integral in the right-hand side of the expression giving $A_2/\kone^2$ in \Eq{eq:Different coefficients expansion small k} is large and we can neglect the boundary term $\tau_\I^2/2$. Besides, noticing that the integral is mainly dominated by the value of the integrand around the time $\tau^*$ at which $z$ becomes small (or even vanishes), which is in the second CR phase, one finds
\bea
\label{eq:A2 approximation}
\frac{A_2}{\kone^2} \simeq -KI
\qquad \mbox{with} \qquad
I=\underline{\mathcal{H}}\left[\int^0_{\tau_\I} \frac{\td \tau'}{z(\tau')^2}\right] \mbox{,} \qquad
K = \int_{-\infty}^{\tau^*}\td \tau'z(\tau')^2 \, .
\eea 
Notice that $K$ remains finite while $I$ diverges when $z$ approaches $0$ at late times. Moreover, since $K$ is always positive, $A_2$ has the opposite sign to $I$. If $z$ never flips sign, $I$ is positive and therefore, $A_2$ is negative. However, if $z$ flips sign, the Hadamard part gives a negative contribution to the integral (see Appendix~\ref{app:Hadamard} for details), hence $I<0$ and $A_2$ is positive.

As discussed in \Sec{sec:intro}, after the possible dip, the spectrum cannot grow faster than $k^4$. Hence, the only relevant terms required to describe it are
\be
\label{eq:Pzeta before peak}
   \mathcal{P}_\zeta(k) =\mathcal{P}_\zeta^{\mathrm{CMB}}\left\{\left[1+A_2\left(\frac{k}{\kone}\right)^2+A_4\left(\frac{k}{\kone}\right)^4\right]^2+B_0^2\left(\frac{k}{\kone}\right)^{6}\right\}\,,
\ee
where we used $\nu_{\I}=3/2$. The term proportional to $B_0^2$ should be included as it allows one to describe the amplitude of the power spectrum at the dip where the contribution in the square brackets vanishes, as explained around \Eq{eq:amplitude dip}. With the approximation \eqref{eq:Pzeta before peak}, the vanishing condition \eqref{eq:condition dip} leads to
\be \label{eq:k_dip_1}
\kdip^2 = -\frac{A_2\kone^2}{2A_4}\left(1\pm\sqrt{1-\frac{4A_4}{A_2^2}}\right)\,,
\ee
where the smallest positive solution should be chosen. In Appendix~\ref{app:A4}, we show that, under the assumption that the power spectrum is strongly amplified, $A_4$ grows as $I$, so $A_2^2\gg A_4$.\footnote{This serves as a consistency check for the $k^4$ growth: due to the largeness of $A_2$, the $A_2$ term in \Eq{eq:Pzeta before peak} dominates even for $k$ values somewhat larger than $k_I$, leading to the $k^4$ behaviour when $\nu_\I = 3/2$. 
} 
This is why $A_4$ can be discarded in \Eq{eq:k_dip_1}, which yields
\be
\label{eq:k_dip}
\kdip^2 = -\frac{\kone^2}{A_2} \, .
\ee
For the dip to exist, we need $\kdip$ to be real, that is, $A_2<0$. As explained above, this is equivalent to the requirement that $z$, hence $\dot\phi$, does not switch sign. We have thus shown the following:
\vspace{2mm}
\begin{mdframed}
\begin{center}
\emph{The power spectrum of the curvature perturbation features a dip if and only if the inflaton's velocity does not flip sign.}
\end{center}
\end{mdframed}
\vspace{2mm}
This criterion, which is one of the main results of this paper, is consistent with the conclusions obtained in~\cite{Wang:2024wxq} in the particular case of a piecewise quadratic potential. 

\subsection{Relationship between the dip and the peak}
\label{sec:dip_vs_peak}

Using \Eqs{eq:Pzeta before peak} and \eqref{eq:k_dip}, we can also compute the amplitude of the dip:
\be
\label{eq:expression amplitude dip}
\mathcal{P}_\zeta^{\mathrm{dip}} = -\frac{\mathcal{P}_\zeta^{\mathrm{CMB}}B_0^2}{A_2^3} \, .
\ee

Above the $k^4$ regime, the large-scale expansion breaks down. This typically happens at a scale $\knp$ where the terms proportional to $A_2$ and $B_0$ in \Eq{eq:Pzeta before peak} are comparable\footnote{Alternatively, $\knp$ may be defined as the scale at which the terms proportional to $A_2$ and $A_4$ are comparable, but this would give the same typical scale.}, that is,
\be
\knp^2 = \frac{A_2^2\kone^2}{B_0^2} \, .
\ee
Since the $k^4$ regime ends close to the peak, we can approximate $\knp\approx \kpeak$, hence the power spectrum at the peak is of order
\be
\mathcal{P}_\zeta^{\mathrm{peak}} \approx \frac{\mathcal{P}_\zeta^{\mathrm{CMB}}A_2^6}{B_0^4} \, .
\ee
Combined with \Eq{eq:expression amplitude dip}, this leads to a universal relationship between the peak and the dip of the power spectrum, namely,
\vspace{2mm}
\begin{mdframed}
\be
\label{eq:scaling dip peak}
\frac{\mathcal{P}_\zeta^{\mathrm{peak}}}{\mathcal{P}_\zeta^{\mathrm{CMB}}} \propto \left(\frac{\mathcal{P}_\zeta^{\mathrm{dip}}}{\mathcal{P}_\zeta^{\mathrm{CMB}}}\right)^{-2} \, .
\ee
\end{mdframed}
\vspace{2mm}
This is the other main result of the present work.
It implies that the height of the peak grows quadratically with the depth of the dip, hence the large-peak condition under which this conclusion was obtained is equivalent to a deep-dip condition. When this condition is not met, we expect slight deviations from \Eq{eq:scaling dip peak}. These are further analysed in~\Sec{sec:smooth_tanh} with an explicit example, where we find that \Eq{eq:scaling dip peak} is remarkably accurate even in cases of modest peaks and dips.

\subsection{Understanding the dip}

In \Sec{sec:dip_or_no_dip}, we discussed the existence of a dip in qualitative terms. Having confirmed that the dip only exists if $z$ does not change sign, let us return to the mechanical analogy of \Sec{sec:Modulus:MSvariable} to gain further insight into this result.

First, let us comment on the $k$-dependence around the transition. As can be confirmed from numerical examples, the transition kicks the mode functions $u_k$ onto trajectories that approach $|u_k|=0$ faster for larger $k$. This is consistent with the fact that the $k^2$ term in the Mukhanov--Sasaki equation is an attractive force, pulling $u_k$ towards the origin.

Then, let us consider the different $z$ behaviours. When $z$ crosses zero after the transition, $u_k$ tends to follow $z$ at small $k$, but cannot go all the way to zero due to the conserved angular momentum discussed in \Sec{sec:Modulus:MSvariable}. The velocity of $u_k$ tends to be large as its modulus decreases, corresponding to a large derivative of $z$; in practice, $u_k$ shoots past the origin on an almost straight line in the complex plane, shifted slightly to respect angular momentum conservation. When $k$ changes, the trajectory stays almost the same, but higher $k$ modes travel along it faster due to the initial boost gained from the transition. At late time, modes with higher $k$ always have a higher amplitude; the power spectrum grows monotonously, with no dip.

When $z$ does not cross zero, there are various possible trajectories. At small $k$, $u_k$ first decreases quickly but then turns around almost 180$^\circ$ in the complex plane and starts to grow, mimicking $z$. However, if $k$ is large enough, the boost from the transition takes $u_k$ all the way close to zero and to the other side, along a similar trajectory as in the sign-flipping case. In between these two extremes, $u_k$ has just the right velocity to balance near the origin for a prolonged period of time, after which it may be deflected to various angles. For these modes, the value of $|u_k|$ is relatively small at late times, since its growth is delayed by the balancing behaviour, leading to a dip in the power spectrum.

Above, we assumed that the initial velocity of $u_k$ determines the behaviours of the modes with respect to each other. With the time-dependent force term $u_k z''/z$ in the Mukhanov--Sasaki equation, we should elaborate on this point. We note the force always repels $u_k$ away from the origin. When the magnitude of $u_k$ is decreasing, the force slows the velocity down. Modes with larger $|u_k|$ are slowed down more, so they can never catch up to modes with smaller $|u_k|$. When the magnitude is increasing, the force increases the velocity. Modes with larger $|u_k|$ are accelerated faster, so again, modes with small $|u_k|$ can never catch up. This guarantees that the hierarchy set by the initial velocity is preserved: modes with a higher $k$ end up travelling further in the original direction of $u_k$.

\section{Analytical insights from an instantaneous transition}
\label{sec:instantaneous_transitions}

Let us now illustrate the above findings with a few example models. Most scenarios producing peaks in the power spectrum feature a rapid transition from an initial CR stage to a USR phase. 
In the limit where the transition is instantaneous, the power spectrum can be computed exactly~\cite{Karam:2022nym}. Let us discuss this limiting case in light of the transfer-matrix formalism.

The background of an instantaneous transition happening at $\tau=\tau_0$ is described by
\be
\label{eq:instantaneous transition}
    \frac{z''}{z} 
    = \frac{\mathcal{A}}{\tau_0}\delta(\tau-\tau_0) +
    \frac{1}{\tau^2}\left\{\begin{array}{ll}
    \nu_\I^2-\frac{1}{4}\, ,   & \quad \tau<\tau_0 \\
    \nu_\II^2-\frac{1}{4} ,    & \quad \tau>\tau_0
    \end{array}\right.
    \,,
\ee
where $\tau_0$ is the time of the transition and we work in the quasi de-Sitter limit where $aH =-1/\tau$.
Imposing continuity for $z$ leads to
\be
\label{eq:z Dirac}
z(\tau) =
    \left\{\begin{array}{ll}
    \zeta_1\left(\frac{\tau_0}{\tau}\right)^{\nu_\I-\frac{1}{2}}\, ,                  & \tau<\tau_0 \\
    \zeta_2\left(\frac{\tau_0}{\tau}\right)^{\nu_\II-\frac{1}{2}} + \left(\zeta_1-\zeta_2\right)\left(\frac{\tau_0}{\tau}\right)^{-\nu_\II-\frac{1}{2}} ,    & \tau>\tau_0
    \end{array}\right.
    \,,
\ee
where we take $\zeta_1 > 0$ without loss of generality; $\zeta_2$ may be positive or negative, determining the sign of $z$ at late times. Here $\zeta_2$ is related to $\mathcal{A}$ according to $\mathcal{A} = \nu_\I + \left(1- 2\zeta_2/\zeta_1\right)\nu_\II$. A strong enhancement of the power spectrum in phase $\II$ requires a long USR period, corresponding to $|\zeta_2| \ll \zeta_1$, indicating some tuning of $\mathcal{A}$.

The transition \eqref{eq:instantaneous transition} gives a particularly simple, $k$-independent transfer matrix~\eqref{eq:MS_phases_tildeT} that connects the two constant-roll phases at time $\tau_0$,
\be\label{eq:transfer_inst}
    T(k) = 
    \begin{pmatrix}
        1 & 0 \\
        \mathcal{A}/\tau_0 & 1
    \end{pmatrix}\,.
\ee
The transfer matrix connecting either the Bessel or Hankel modes in the CR phases is obtained using Eqs.~\eqref{eq:TJ} and~\eqref{eq:TH}, respectively. The power spectrum~\eqref{eq:Pzeta:gen} can then be obtained from~\Eq{eq:a_II_in_T}. It reads
\bea
\label{eq:Pzeta_dirac}
   \mathcal{P}_\zeta(k) 
    &= \mathcal{P}^{\rm (CR)}_\zeta(k;\nu_\II) \left(\frac{\pi}{2}\right)^2\bigg|
   (-\mathcal{A}+\nu_\I+\nu_\II) J_{\nu_\II}(-k\tau_0)H_{\nu_\I}(-k\tau_0) \\
    &\qquad\qquad\qquad\qquad + k\tau_0(J_{\nu_\II}(-k\tau_0)H_{\nu_\I-1}(-k\tau_0)
    + J_{\nu_\II+1}(-k\tau_0)H_{\nu_\I}(-k\tau_0))\bigg|^2
    \,.
\eea
This matches the expression given in~\cite{Karam:2022nym}.

\subsection{Features at small scales}
One can expand \Eq{eq:Pzeta_dirac} when $k$ is large to get
\be
\label{eq:Instantaneous:Largek:Pzeta}
   \mathcal{P}_\zeta(k) = \mathcal{P}^{\rm (CR)}_\zeta(k;\nu_\II)\left[1+\mathcal{A}\frac{\cos\left(2k\tau_0+\pi\nu_\II\right)}{k\tau_0}\right] .
\ee
\begin{figure}[t]
\centering
\includegraphics[width=0.47\textwidth]{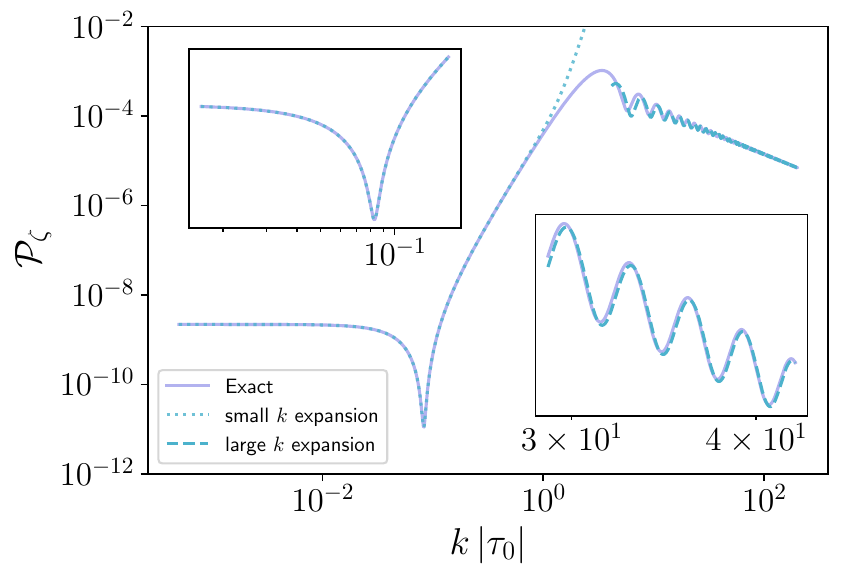}
\includegraphics[width=0.49\textwidth]{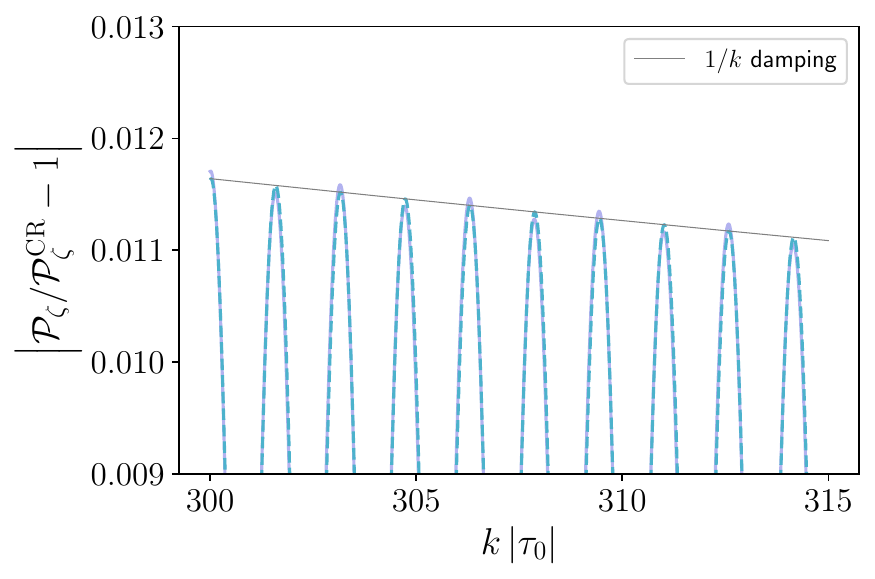}
\caption{\emph{Left panel:} Power spectrum for the instantaneous transition model~\eqref{eq:instantaneous transition}. We have chosen $\nu_\I = 3/2$ and  $\nu_\II = 2$, $\zeta_1$ is set so that the power spectrum at low $k$ matches the amplitude measured by Planck and $\zeta_2$ is such that the maximal value of the power spectrum is $10^{-3}$. The dotted line at small $k$ corresponds to the large-scale expansion~\eqref{eq:expansion instantaneous small k} while the dashed line at large $k$ is the small-scale expansion~\eqref{eq:Instantaneous:Largek:Pzeta}, where $\mathcal{P}^{\mathrm{CR}}_\zeta(k)\propto k^{3-2\nu_\II}$, see \Eq{eq:CR Pzeta}. \emph{Right panel:} Damping of the oscillations at large $k$. The envelope of oscillations as predicted by~\Eq{eq:Instantaneous:Largek:Pzeta} is displayed in thin grey and corresponds to $k^{-1}$ damping.}
\label{fig:bound instantaneous transition}
\end{figure}
This modulation perfectly agrees with the prediction of~\Eq{eq:Pzeta large k first correction} derived using the transfer-matrix formalism when the background is described by~\Eq{eq:instantaneous transition}. In this case, there are no boundary terms since $\tau_\I=\tau_\II$. In~\Fig{fig:bound instantaneous transition}, the power spectrum~\eqref{eq:Pzeta_dirac} is compared with the small-scale approximation~\eqref{eq:Instantaneous:Largek:Pzeta} and one can check that the agreement is excellent. The correction to the power spectrum oscillates at frequency $2\tau_0$, and the envelope of the oscillations decays as $k^{-1}$, in agreement with the considerations presented in~\Sec{sec:sub_hubble_power_spectrum}. Note that, in practice, the instantaneous transition can only be an approximation, and transitions in physical models always occur over a finite duration $\Delta\tau$, even if short. When $k$ is large enough to ``resolve'' the transition, \ie~$k\Delta\tau\gg 1$, non-perturbative decay is recovered.

\subsection{Features at large scales}

The criterion for the existence of a dip in the power spectrum and the relation between the amplitude of this dip and the peak of the power spectrum predicted above can be checked explicitly in this particular model of an instantaneous transition between two constant-$\nu$ phases. At large scales, the expansion of the power spectrum~\eqref{eq:Pzeta_dirac} at next to leading order in $k$ gives
\be
\label{eq:expansion instantaneous small k}
    \mathcal{P}_\zeta(k) = \mathcal{P}^{\rm (CR)}_\zeta(k,\nu_\I)\left\{1+\frac{k^2\tau_0^2}{4\left(\nu_\I-1\right)\left(\nu_\II+1\right)}\left[2+\nu_\II-\nu_\I-\frac{\zeta_1}{\zeta_2}\left(1+\frac{\nu_\I}{\nu_\II}\right)\right]\right\}^{2}\,.
\ee
This result can be compared with the expansion of the power spectrum obtained in \Eq{eq:Pzeta expansion small k}. For an instantaneous transition, the coefficient $A_2$ can be computed using \Eq{eq:Different coefficients expansion small k} and the expression for the function $z(\tau)$ derived in \Eq{eq:z Dirac}. This yields
\be
\label{eq:A2 instantaneous}
    \frac{A_2}{k_0^2} = \frac{\tau_0^2}{4\left(\nu_\I-1\right)\left(\nu_\II+1\right)}\left[2+\nu_\II-\nu_\I-\frac{\zeta_1}{\zeta_2}\left(1+\frac{\nu_\I}{\nu_\II}\right)\right]\,,
\ee
where $k_0 = -1/\tau_0$ (in the instantaneous case, $\tau_0$ plays the role of both $\tau_\I$ and $\tau_\II$). This result, together with \Eq{eq:Pzeta expansion small k}, perfectly agrees with the expansion \eqref{eq:expansion instantaneous small k}. According to the discussion below \Eq{eq:k_dip}, if the power spectrum is strongly enhanced, that is when the ratio $\zeta_1/|\zeta_2|$ is large, a dip exists if $A_2<0$, \ie~typically when $\zeta_1$ and $\zeta_2$ have the same sign. Notice that in this case, $z$ does not flip sign, which is consistent with the discussion in~\Sec{sec:dip_crit}.

We would now like to check that the relation between the amplitude of the peak and the one of the dip, \Eq{eq:scaling dip peak}, is indeed satisfied.
We first have to compute the coefficient $B_0$. Using again \Eq{eq:Different coefficients expansion small k} and the function $z(\tau)$ in \Eq{eq:z Dirac}, we get
\be
\label{eq:B0 instantaneous}
    \frac{B_0}{\kone^{2\nu_\I}} = \left(\frac{-\tau_0}{2}\right)^{2\nu_\I}\frac{\Gamma(1-\nu_\I)}{\Gamma(1+\nu_\I)}\left(\frac{\nu_\I\zeta_1}{\nu_\II\zeta_2}-1\right)\,.
\ee
Recall that \Eq{eq:scaling dip peak} was derived under the assumption that $\nu_\I=3/2$ and that the power spectrum features a large peak. The latter condition corresponds to $\left|\zeta_2\right| \ll \zeta_1$, as discussed above. Under these assumptions,
\be
\label{eq:Pzeta instantaneous SR}
\begin{split}
    \mathcal{P}_\zeta(k) = \mathcal{P}_\zeta^{\mathrm{CMB}}&\left|2i\sqrt{\pi}\Gamma\left(1+\nu_\II\right)\left(\frac{-k\tau_0}{2}\right)^{\frac{3}{2}-\nu_\II}J_{\nu_\II}H_{\frac{3}{2}}\right. \\
    &\left.-\frac{2i\sqrt{\pi}\Gamma\left(\nu_\II\right)\zeta_1}{\zeta_2}\left(\frac{-\tau_0k}{2}\right)^{\frac{5}{2}-\nu_\II}\left(J_{\nu_\II}H_{\frac{1}{2}}+J_{\nu_\II+1}H_{\frac{3}{2}}\right)\right|^2\,,
\end{split}
\ee
where we recall that the Bessel and Hankel functions are evaluated at $-k\tau_0$ (we dropped their argument for conciseness). Because $\abs{\zeta_2}\ll\zeta_1$, the power spectrum close to the peak is dominated by the second term inside the absolute value, hence its position is independent of the ratio $\zeta_1/|\zeta_2|$ and its amplitude at the peak reads
\be
\label{eq:Pzeta peak instantaneous}
    \frac{\mathcal{P}_\zeta^{\mathrm{peak}}}{\mathcal{P}_\zeta^{\mathrm{CMB}}} \approx \left|\frac{2i\sqrt{\pi}\Gamma\left(\nu_\II\right)\zeta_1}{\zeta_2}\left(\frac{-\tau_0\kpeak}{2}\right)^{\frac{5}{2}-\nu_\II}\left(J_{\nu_\II}H_{\frac{1}{2}}+J_{\nu_\II+1}H_{\frac{3}{2}}\right)\right|^2\,.
\ee
Thus, we obtain that the power spectrum at the peak scales as $\left(\zeta_1/\zeta_2\right)^2$. On the other hand, the dip, if it exists, occurs at a small $k$ so that the Bessel and Hankel functions can be Taylor expanded close to $-k\tau_0=0$ in \Eq{eq:Pzeta instantaneous SR}. Keeping terms up to order $k^2$, we find that the dip is located at
\be
\left(-\kdip\tau_0\right)^2 = \frac{4\nu_\II\left(1+\nu_\II\right)\zeta_2}{\left(2\nu_\II+3\right)\zeta_1}\,,
\ee
which is indeed small. The amplitude of the power spectrum at the dip is then
\be
\label{eq:Pzeta dip instantaneous}
\frac{\mathcal{P}_\zeta^{\mathrm{dip}}}{\mathcal{P}_\zeta^{\mathrm{CMB}}} \approx \frac{16\nu_\II\left(1+\nu_\II\right)^3\zeta_2}{\left(2\nu_\II+3\right)^3\zeta_1} \,.
\ee
In particular, we observe that this amplitude scales as $\zeta_2/\zeta_1$. Together with \Eq{eq:Pzeta peak instantaneous}, this confirms that the amplitude of the peak scales as the squared inverse of the amplitude of the dip, as obtained in \Eq{eq:scaling dip peak}.

\section{Smooth transitions and non-perturbative decay}
\label{sec:example_damping}

\subsection{An analytic toy model}
\label{sec:analytic_toy_model}

In order to highlight the non-perturbative nature of the decay of oscillations in the case of a smooth transition, see \Sec{sec:sub_hubble_power_spectrum}, we now consider a toy model that can be solved analytically. The model is defined by a hyperbolic pulse at $\tau_0$, with a duration $\Delta \tau$,
\be\label{eq:zpp_tanh}
    \frac{z''}{z} = \frac{\mathcal{A}}{2\tau_0 \Delta \tau} \left[1-\tanh^2\left(\frac{\tau-\tau_0}{\Delta \tau}\right)\right],
    \qquad
    \tau \in [\tau_\I,\tau_\II] \, .
\ee
The amplitude of the pulse is normalized so that $\tilde A(0) = \mathcal{A}/\tau_0$, see \Eq{eq:A(k):def}. With this ansatz, the Mukhanov--Sasaki equation~\eqref{eq:MS} has two independent solutions $u_{\pm k}\left(\tau\right)$ given by\footnote{The general Legendre equation can be obtained with the change of variables $\tau - \tau_0 = \Delta \tau\, {\mathrm{arctanh}}(x)$.}
\bea
\label{eq:uk:tanh:Legendre}
    u_k(\tau) =& \Gamma(1-ik\Delta\tau)P^{ik\Delta\tau}_{\bar{\mathcal{A}}}\left[-\tanh\left(\frac{\tau - \tau_0}{\Delta \tau}\right)\right]
    \quad\text{with}\quad
    \bar{\mathcal{A}} \equiv \frac{1}{2}\left(\sqrt{1-2 \mathcal{A} \frac{\Delta\tau}{\tau_0}} - 1\right) \,,
\eea
where $P^{\mu}_{\nu}$ denotes the associated Legendre function. This gives rise to an explicit expression for the Wronskian matrix~\eqref{eq:Wronskian}, hence for the transfer matrix $\T=W(\tau_\II)W^{-1}(\tau_\I)$, which in turn leads to an exact expression for the power spectrum using \Eqs{eq:TH} and~\eqref{eq:a_II_in_T}.

The ansatz \eqref{eq:zpp_tanh} describes a transition between two constant-$\nu$ phases with $\nu = 1/2$ in each phase. Such phases would produce a blue-tilted spectrum with $n_{\mathrm{s}} = 3$, which is excluded by CMB measurements at large scales, so another transition would have to be added to make this scenario viable. This possibility is further investigated in~\Sec{sec:smooth_tanh}, but here, we only use \Eq{eq:zpp_tanh} as a toy model to highlight the existence of exponentially decaying oscillations.

Moreover, as we are interested in modes that are mostly sub-Hubble during the transition, $k^{-2}z''/z$ is small away from the pulse, and we can neglect it in the constant-$\nu$ phases as an approximation. This would justify extending $\tau_\I$, $\tau_\II$ to $\pm \infty$ without affecting the result significantly. However, the ansatz would then not capture mode evolution away from the pulse and omit the boundary terms in \Eq{eq:A(k):IntByPart}, which can be relevant for the large-$k$ expansion.

We also remark that the viable parameter range for the ansatz \eqref{eq:zpp_tanh} is limited to 
\be\label{eq:toy_conditions}
    \Delta \tau/|\tau_0| \lesssim 1, 
    \qquad\qquad
    \bar{\mathcal{A}} \lesssim 1.
\ee
As the transition takes place roughly in the range $(\tau_0 - \Delta \tau, \tau_0 + \Delta \tau)$, the first condition imposes that it ends before future infinity. Moreover, as long as $\Delta \tau/|\tau_0| \ll 1$, this ratio gives the duration of the transition in $e$-folds $\Delta \tau/|\tau_0| \sim \Delta N$, so we expect the ansatz to work better for relatively brief transitions. The second condition $\bar{\mathcal{A}} \lesssim 1$ arises if one wishes to avoid oscillations in $z$. The latter would imply oscillations in $\dot \phi = z/a$, which may be difficult to realize consistently in models of single-field inflation (see \Sec{sec:reconstructing_V} for a related discussion). Since $z$ is given by a $k=0$ solution of the Mukhanov--Sasaki equation, it can be described by a Legendre function like the rest of the modes~\eqref{eq:uk:tanh:Legendre}. For $\bar{\mathcal{A}} \geq 1$, the Legendre functions can have multiple zeros, so $z$ can flip sign multiple times. Note, however, that even for $\bar{\mathcal{A}} \in (0,1]$, the Legendre functions have a zero and thus $z$ flips its sign as it evolves.

\subsubsection{Transfer matrix and the power spectrum}

The power spectrum can be computed explicitly at late times by using the asymptotics of $u_{k}$ from \Eq{eq:uk:tanh:Legendre},
\be\label{eq:toy_asymptotes}
    u_k(\tau) = \left\{\begin{array}{ll}
        e^{-ik(\tau-\tau_0)} &, \quad \tau_0-\tau \gg \Delta\tau 
        \\
        e^{-i k (\tau-\tau_0) } \alpha_k 
        + e^{i k (\tau-\tau_0) } \beta_k 
         &, \quad \tau-\tau_0 \gg \Delta\tau 
    \end{array}\right. \,,
\ee
where
\be
    \alpha_k = \frac{
        \Gamma (1-i k \Delta \tau) \Gamma (-i k \Delta \tau)}{\Gamma \left(1 -i k \Delta \tau 
 +\bar{\mathcal{A}} \right) \Gamma \left(-i k\Delta \tau - \bar{\mathcal{A}} \right)  }\, ,
    \qquad
    \beta_k = i \frac{\sin (\pi \bar{\mathcal{A}})}{\sinh(\pi  k \Delta \tau)}\, .
\ee
Since $u_k$ and $u_k^*=u_{-k}$ are two independent modes, the conservation of their Wronskian determinant implies that $|\alpha_k|^2 -  |\beta_k|^2 = 1$ as is expected for a Bogoliubov transformation. In particular, since $\beta_k$ is exponentially suppressed for large $k$, $\alpha_k$ must approach 1 exponentially when $k\Delta \tau \gg 1$. For completeness, note that since the asymptotes in \Eq{eq:toy_asymptotes} are already reported in the Hankel basis (for $\nu_i = 1/2$ we have $\tilde{u}_i^{(1,2)}\propto e^{\mp i k \tau}$), the transfer matrix can be written as
\be
    T_H(k) =
    \begin{pmatrix}
        \alpha_k & \beta_k \\
        \beta_k^{*} & \alpha_k^{*}
    \end{pmatrix}
\ee
and has an explicit U(1,1) structure.

\begin{figure}[t]
\centering
\includegraphics[width=0.48\textwidth]{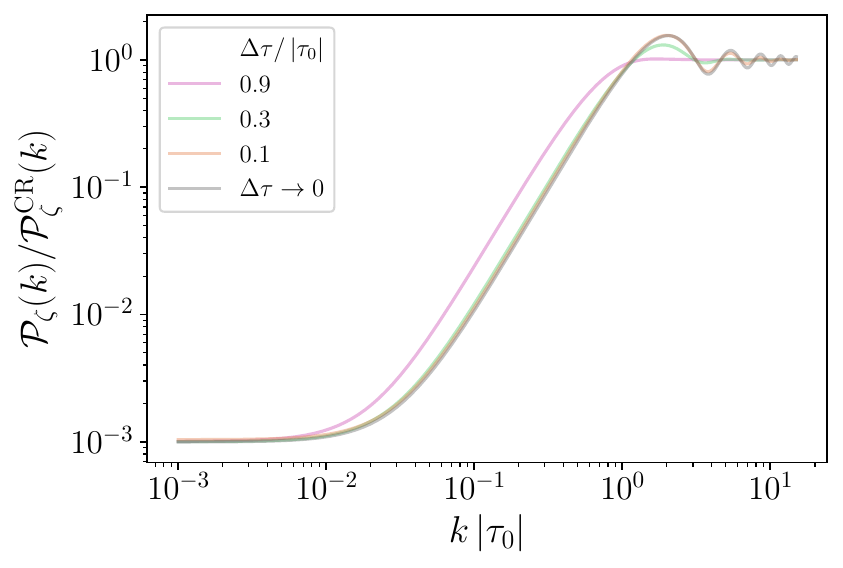}
\includegraphics[width=0.49\textwidth]{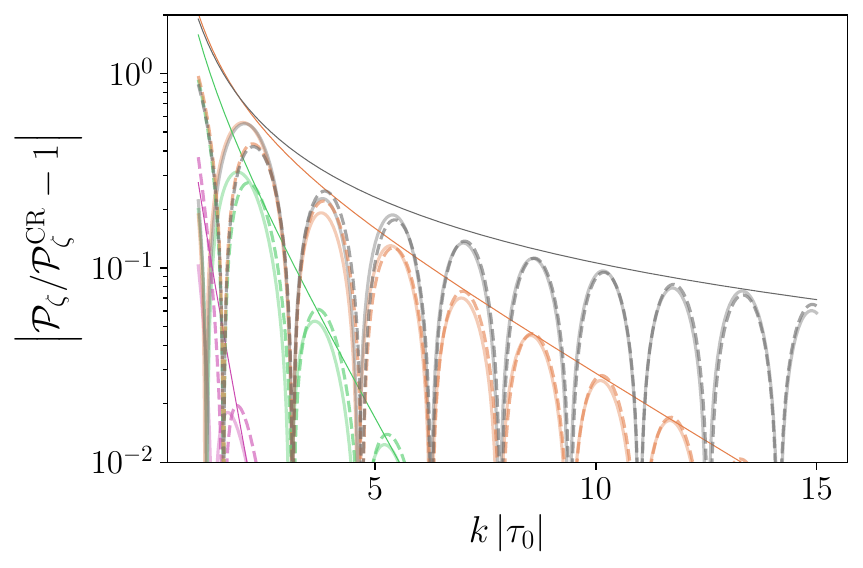}
\caption{\emph{Left panel:} Curvature power spectra of the hyperbolic-pulse model~\eqref{eq:zpp_tanh} for different values of $\Delta \tau/|\tau_0|$, normalised by its constant-roll counterpart. The instantaneous limit $\Delta \tau \to 0$ is shown in grey. The spectra are normalized to the asymptotic $\mathcal{P}^{\rm (CR)}_\zeta(k;\nu_\II) \propto k^2$ power spectrum. The values of $\bar{\mathcal{A}}$ are chosen in a way to produce a $10^3$ enhancement after the transition, which leads to the following amplitudes $\bar{\mathcal{A}} = (0.3928,~0.1536,~0.0516)$ in the order of decreasing $\Delta \tau/|\tau_0|$. \emph{Right panel:} The damping of spectral oscillations at small scales. The thin solid lines show the approximate envelope of spectral oscillations given by \Eq{eq:spectral_bound} and the dashed lines display the large-$k$ expansion~\eqref{eq:tanh:model:Pzeta:small:scale}.}
\label{fig:bound_toy_model}
\end{figure}

For a Bunch--Davies initial state, this gives rise to the power spectrum 
\bea
\label{eq:tanh:model:Pzeta:small:scale:exact}
  { \mathcal{P}_\zeta(k)}
  &\ \ \, = \ \ \, \mathcal{P}^{\mathrm{ (CR)}}_\zeta(k;\nu_\II) \left[
   1 + 2 \Re\left(\beta_k \alpha_k^{*}e^{-i2k\tau_0}\right) + 2|\beta_k|^2\right] 
     \\
  &\stackrel{k\Delta\tau \gg 1}{\simeq} \mathcal{P}^{\mathrm{ (CR)}}_\zeta(k;\nu_\II) \left[
   1 + 2 \sin\left(\pi \bar{\mathcal{A}}\right) \frac{\sin\left(2k\tau_0\right)}{\sinh(\pi k \Delta \tau)} \right]
     \, ,
\eea
where $\nu_\II = 1/2$ and we have assumed $|k\tau|\ll 1$. The second line gives the leading exponentially-suppressed terms, that is, we took $\alpha_k \sim 1$ and dropped the $|\beta_k|^2$ term. We also omitted a $k^{-1}$ suppressed phase. The first line is exact and it implies the upper bound\footnote{The triangle inequality gives $(|\alpha_k|-|\beta_k|)^2 \leq \mathcal{P}_\zeta(k)/\mathcal{P}^{\mathrm{ (CR)}}_\zeta(k;\nu_\II) \leq (|\alpha_k|+|\beta_k|)^2$, which, by $|\alpha_k|^2 = |\beta_k|^2 + 1$, reduces to $-2|\alpha_k \beta_k|+2|\beta_k|^2\leq \mathcal{P}_\zeta(k)/\mathcal{P}^{\mathrm{ (CR)}}_\zeta(k;\nu_\II) - 1 \leq 2|\alpha_k \beta_k|+2|\beta_k|^2$. This is slightly stronger than \Eq{eq:spectral_bound}, especially for downward fluctuations.}
\be\label{eq:spectral_bound}
    \left| \mathcal{P}_\zeta(k)/\mathcal{P}^{\mathrm{ (CR)}}_\zeta(k;\nu_\II) - 1 \right| \leq 2|\beta_k|+4|\beta_k|^2
\ee
that provides a good fit to the envelope of the oscillations, see \Fig{fig:bound_toy_model}.
Interestingly, a non-perturbative property of the ansatz \eqref{eq:zpp_tanh} is that $\beta_k = 0$ when $\bar{\mathcal{A}} $ is an integer hence all spectral modulations vanish periodically in $\bar{\mathcal{A}}$. However, as $\bar{\mathcal{A}} \gtrsim 1$ is incompatible with \Eq{eq:toy_conditions}, it is not clear whether such scenarios can be realized in single-field models.

A dip at low-$k$ is seen to be absent in the spectra displayed in \Fig{fig:bound_toy_model}. This is consistent with the criteria of \Sec{sec:suphorizon} because, as discussed above, $z$ must flip its sign for the values of $\bar{\mathcal{A}}$ in~\Fig{fig:bound_toy_model}, which are required to obtain a large spectral enhancement.

\subsubsection{Small-scale limit} 

Let us now compare this result with the one obtained from the large-$k$ expansion scheme presented in \Sec{sec:subH:exp}. Assuming that $\tau_0-\tau_\I, \tau_\II-\tau_0 \gg \Delta\tau$, one can show that
\be
    \tilde A(k) = \frac{\pi \mathcal{A} k \Delta \tau}{2\sinh(\pi k \Delta \tau/2)} e^{ik\tau_0}\, .
\ee
In this same limit, the contribution from the boundary $\tilde{B}$ terms vanishes, and with $\nu_\II=1/2$, \Eq{eq:Pzeta large k first correction} reduces to 
\bea
\label{eq:tanh:model:Pzeta:small:scale}
  {\mathcal{P}_\zeta(k)}&= \mathcal{P}^{\mathrm{ (CR)}}_\zeta(k;\nu_\II) \left[
   1 - \frac{\pi \mathcal{A}   \Delta \tau/\tau_0}{\sinh(\pi k \Delta \tau)} \sin\left(2k\tau_0\right)\right]
     \, .
\eea
This coincides with \Eq{eq:tanh:model:Pzeta:small:scale:exact} only in the limit $\mathcal{A}\Delta\tau/|\tau_0|\ll 1$, where $2 \sin(\pi \bar{\mathcal{A}})\simeq -\pi \mathcal{A}\Delta\tau/\tau_0$. This is because \Eq{eq:tanh:model:Pzeta:small:scale} was derived using the perturbative scheme described in \Sec{sec:subH:exp}, which only captures contributions linear in $\mathcal{A}$. Even though this approximation correctly describes the frequency and the phase of the oscillations, as well as the way they are damped with $k$, their overall amplitude is properly estimated only in the regime $\mathcal{A}\Delta\tau/|\tau_0|\ll 1$. This highlights the fact that the framework of \Sec{sec:subH:exp} is not only a large-scale expansion, it is also an expansion in the amplitude and duration of the pulse.
The comparison between the full power spectrum and the small-scale approximation~\eqref{eq:tanh:model:Pzeta:small:scale} is displayed in~\Fig{fig:bound_toy_model}, where one can check that the agreement is excellent at large $k$ if $\mathcal{A}\Delta\tau/|\tau_0|\ll 1$. 

It also follows from~\Eq{eq:tanh:model:Pzeta:small:scale:exact} or~\Eq{eq:tanh:model:Pzeta:small:scale} that the spectral oscillations proceed at frequency $2\tau_0$. When $k\Delta\tau\ll 1$, the amplitude of the oscillations is controlled by $\mathcal{A}/k$, while it is suppressed by $e^{-\pi k \Delta\tau}$ when $k \Delta\tau\gg 1$. This is consistent with the discussion at the end of \Sec{sec:sub_hubble_power_spectrum}. 

\subsubsection{Instantaneous  limit} 

Let us now consider the instantaneous limit where $\Delta \tau/|\tau_0| \ll 1$. In this regime, we have
\be
    \alpha_k 
    =1 - \beta_k ,
    \qquad
    \beta_k = -\frac{i \mathcal{A}}{2k \tau_0}\left(1 - \mathcal{A} \frac{\Delta\tau}{\tau_0}\right) + \mathcal{O}(\Delta\tau^2)\,,
\ee
and thus the power spectrum reads
\bea
\label{eq:tanh:model:Pzeta:small:scale:exact:inst:limit}
  {\mathcal{P}_\zeta(k)}
  & = \mathcal{P}^{\mathrm{ (CR)}}_\zeta(k;\nu_\II) \left\{
   1\!-\!\mathcal{A}\left(1\!-\!\mathcal{A} \frac{\Delta\tau}{\tau_0}\right)
   \frac{\sin(2 k \tau_0)}{k \tau_0} +  \left[\mathcal{A}\left(1\!-\!\mathcal{A} \frac{\Delta\tau}{\tau_0}\right) 
   \frac{\sin(k \tau_0)}{k \tau_0} \right]^2\right\} \,.
\eea
The exponential suppression of spectral oscillations at large $k$ is replaced by a slow $k^{-1}$ decay, characteristic of instantaneous transitions. Indeed, the power spectrum \eqref{eq:tanh:model:Pzeta:small:scale:exact:inst:limit} matches exactly the one of instantaneous transitions \eqref{eq:Pzeta_dirac} with $\nu_\I = \nu_\II = 1/2$ when one replaces $\mathcal{A}$ in Eq.~\eqref{eq:tanh:model:Pzeta:small:scale:exact:inst:limit} with the effective strength of the feature $\mathcal{A}_{\rm eff} \equiv \mathcal{A}\left(1 - \mathcal{A} \Delta\tau/\tau_0\right)$. Thus, the leading term in $\Delta\tau$ can be interpreted as a non-linear correction when matching a sharp but finite transition to an instantaneous one. Therefore, at leading order in $\Delta\tau$, the damping of spectral oscillations is effectively not modified with respect to the instantaneous case. \Fig{fig:bound_toy_model} confirms that the instantaneous limit gives an excellent approximation for the first spectral oscillations when $\Delta\tau/\tau_0 \lesssim 0.1$.

\subsection{Smooth transitions for general constant-$\nu$ phases}
\label{sec:smooth_tanh}

To test the above results in a more generic setting, let us consider a smooth ansatz that interpolates between constant-$\nu$ phases with a strongly time-dependent feature in the transition:
\bea
\label{eq:smooth tanh}
    \frac{z''}{z} 
    =& \frac{1}{\tau^2}
    \left\{ \nu_\I^2 + \frac{\nu_\II^2-\nu_\I^2}{2}\left[1 + \tanh\left(\frac{\tau-\tau_0}{\Delta\tau}\right)\right] - \frac{1}{4}\right\} 
    + \frac{A}{2\tau_0\Delta\tau}\left[1-\tanh\left(\frac{\tau-\tau_0}{\Delta\tau}\right)^2\right] \,.
\eea
The first line represents a smooth and gradual change in $\nu$, while the second line describes a feature with amplitude $A$ which needs to be tuned to obtain an enhancement of the power spectrum. Unlike the ansatz \eqref{eq:zpp_tanh}, this model cannot be solved analytically, and we need to resort to numerical methods to find accurate solutions.

\begin{figure}[t]
\centering
\includegraphics[width=0.48\textwidth]{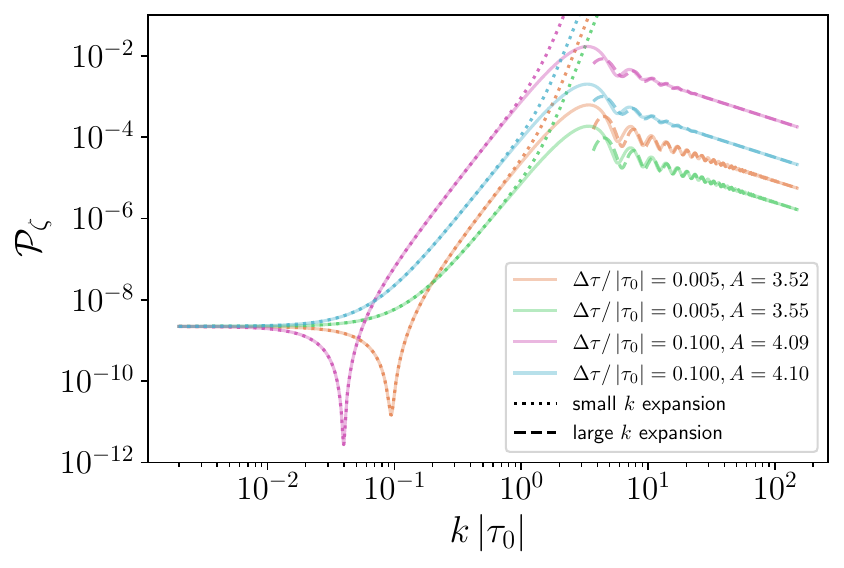}
\includegraphics[width=0.49\textwidth]{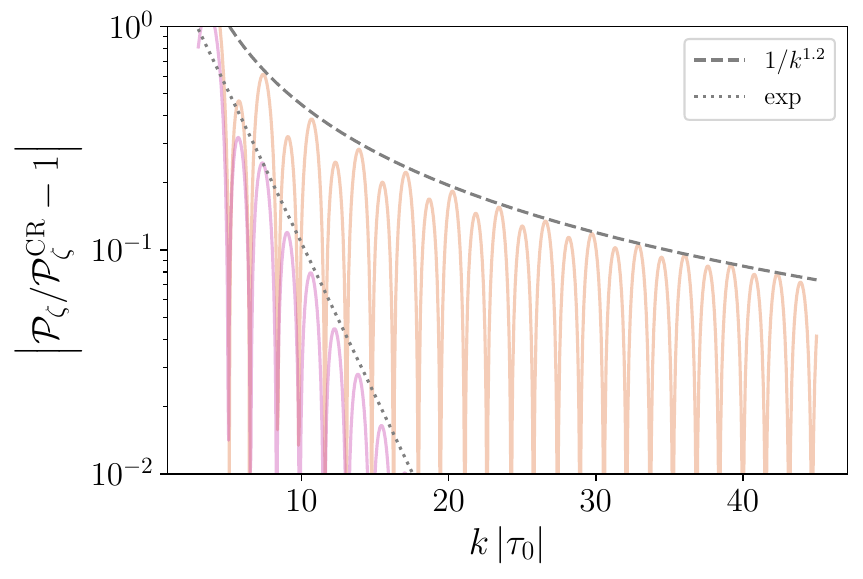}
\caption{\emph{Left panel:} Curvature power spectra of the smooth model~\eqref{eq:smooth tanh} for different values of $\Delta \tau/|\tau_0|$. The spectra are normalized at small $k$ to the value measured by Planck. We have chosen $\nu_\I=3/2$ and $\nu_\II=2$. \emph{Right panel:} Damping of spectral oscillations. The orange solid line shows a power-law damping while the purple one exhibits exponentially-damped oscillations.}
\label{fig:power spectrum and damping tanh}
\end{figure}

The comparison between the exact power spectrum, obtained numerically, and the small-scale approximation~\eqref{eq:Pzeta large k first correction} as well as the large-scale expansion~\eqref{eq:Pzeta before peak} is displayed in the left panel of \Fig{fig:power spectrum and damping tanh}. This confirms that at large  scales, the power spectrum and its possible dip are perfectly described by the coefficients $A_2$, $A_4$, and $B_0$.
In the right panel, we zoom in on the damped oscillations at large $k$. We find two different kinds of damping. When the transition is sharp ($\Delta\tau/\tau_0 = 0.005$), the damping follows a power law while when the transition is smooth ($\Delta\tau/\tau_0 = 0.1$), oscillations decay exponentially. This is consistent with the discussion in \Sec{sec:sub_hubble_power_spectrum}. Notice that the power-law exponent we fit is not an integer probably because both $1/k$ and a $1/k^2$ contributions are relevant in this range of scales.

\begin{figure}[t]
\centering
\includegraphics[height=0.32\textwidth,trim={0cm -0.2cm 0cm 0cm}, clip]{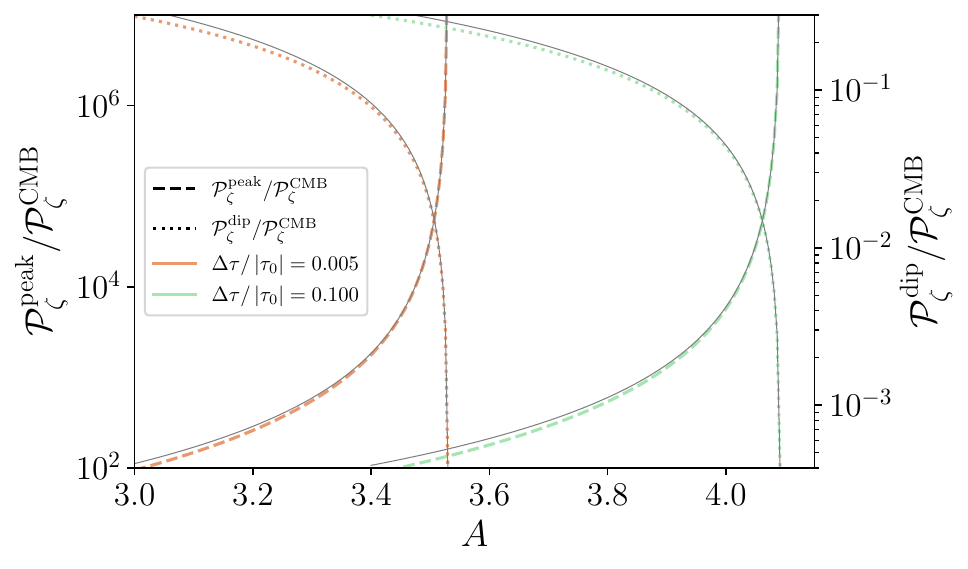}
\includegraphics[height=0.32\textwidth,trim={0cm 0cm 0cm 0cm}, clip]{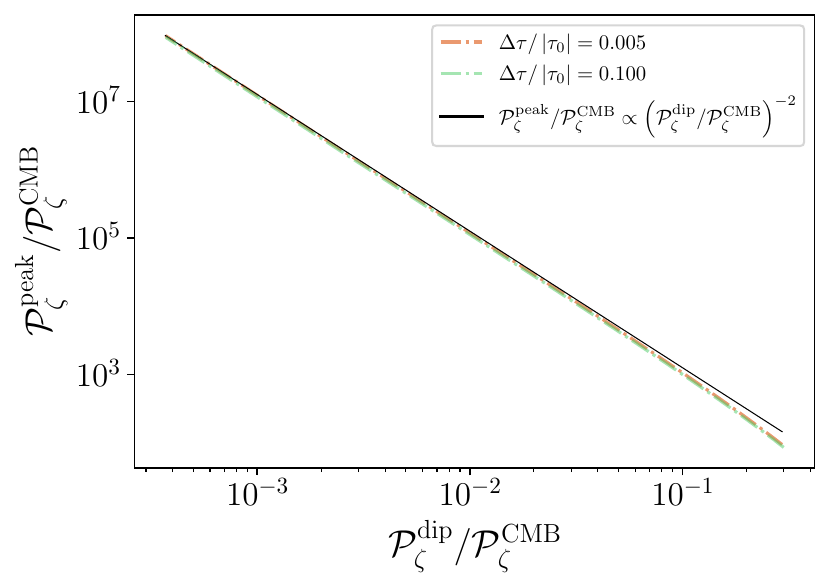}
\caption{\emph{Left panel:} Amplitudes of the peak and of the dip as a function of $A$ for two different values of $\Delta\tau/\tau_0$. We have chosen $\nu_\I=3/2$ and $\nu_\II=2$. The thin gray lines are a fit to the curves and correspond to $\mathcal{P}_\zeta^{\mathrm{peak}}\propto \left(1-A/A_{\mathrm{c}}\right)^{-2}$ and $\mathcal{P}_\zeta^{\mathrm{dip}}\propto \left(1-A/A_{\mathrm{c}}\right)$ where $A_{\mathrm{c}}$ is the value in parameter space at which the power spectrum diverges. \emph{Right panel:} Amplitude of the peak as a function of the amplitude of the dip for two different values of $\Delta\tau/\tau_0$. The black solid line corresponds to the relation derived in~\Sec{sec:dip_vs_peak}.}
\label{fig:dip and scaling dip tanh}
\end{figure}

We also varied the value of the parameter $A$ in this smooth model to obtain different power spectra with various amplitudes for the dip and the peak. In the left panel of Fig.~\ref{fig:dip and scaling dip tanh}, we display the amplitude of the peak and that of the dip as a function of the amplitude of the feature $A$ for two different values of the ratio $\Delta\tau/\tau_0$. The thin gray lines are a fit to the curves and they correspond to $\mathcal{P}_\zeta^{\mathrm{peak}}\propto \left(1-A/A_{\mathrm{c}}\right)^{-2}$ and $\mathcal{P}_\zeta^{\mathrm{dip}}\propto \left(1-A/A_{\mathrm{c}}\right)$ where $A_{\mathrm{c}}$ is the value in parameter space at which the power spectrum diverges. In the right panel, we show  these two amplitudes on a logarithmic scale. In the limit of strong enhancement, that is, when the ratio $\mathcal{P}_\zeta^{\mathrm{peak}}/\mathcal{P}_\zeta^{\mathrm{CMB}}$ is large, we recover the relation~\eqref{eq:scaling dip peak}, to which only small deviations are observed when the power spectra are enhanced by less than two orders of magnitude. This confirms the robustness of our analysis.

Importantly, the same relationship between the dip and the peak is found for different values of $\Delta \tau/\tau_0$, as shown in \Fig{fig:dip and scaling dip tanh}. This suggests that not only the scaling is universal, but the numerical prefactor appearing in \Eq{eq:scaling dip peak} might also be somewhat model-independent. We leave the investigation of this possibility to future work.

\subsection{On reconstructing the potential from an ansatz for background evolution}
\label{sec:reconstructing_V}

The behaviours of $z''/z$ considered in this work are fairly common for inflationary models considered in the literature (see e.g.~\cite{Karam:2022nym} for discussion and examples). Nevertheless, when starting from $z''/z$, as we did in~\Sec{sec:analytic_toy_model}, it is fair to ask which inflationary potential $V(\phi)$ (if any) produces this behaviour, assuming canonical single-field inflation. To reconstruct $V(\phi)$ from $z''/z$, let us first choose some initial values for $a$, $\mathcal{H}$, and $z$. Then, $z(\tau)$ can be obtained as the growing solution from the fixed form of $z''/z$, corresponding to an inflationary attractor at early and late times. By using the Klein-Gordon and Friedmann equations in conformal time,
\begin{equation} \label{eq:Friedmann_conformal}
    \phi'' + 2\mathcal{H}\phi' + a^2V = 0 \, , \quad
    3\mathcal{H}^2 = \frac{1}{2}\phi'^2 + a^2V \, ,
\end{equation}
we can derive
\begin{equation} \label{eq:H_equation}
    \mathcal{H}' - \mathcal{H}^2 = -\frac{1}{2}\phi'^2 = -\frac{\mathcal{H}^2}{2a^2}z^2 \, .
\end{equation}
This allows one to solve for $\mathcal{H}(\tau)$ given the known $z(\tau)$ function and initial conditions, hence to reconstruct $a(\tau)$. Then $z=a\phi'/\mathcal{H}$ also gives $\phi'(\tau)$ and thus $\phi(\tau)$, and the Friedmann equation~\eqref{eq:Friedmann_conformal} can be used to derive $V(\tau)$. Finally, upon inverting the $\phi(\tau)$ function, the potential function $V(\phi)$ can be obtained. There is some freedom in the solution related to the initial conditions, which set the energy scale of inflation in particular.

In principle, this construction can be carried out for any $z''/z$. If we demand the resulting potential function to be positive, \Eq{eq:Friedmann_conformal} gives the consistency condition
\begin{equation} \label{eq:z_max}
    z^2 < 6a^2 \, .
\end{equation}
This corresponds to the condition $\epsilon_H < 3$, where $\epsilon_H$ is the first slow-roll parameter; the condition also easily follows from the familiar form
\begin{equation}
    \epsilon_H = \frac{\dot{\phi}^2}{2H^2} = \frac{3\dot{\phi}^2}{\dot{\phi}^2 + 2V} \, .
\end{equation}
Condition \eqref{eq:z_max} is not very restrictive, and it can be satisfied by tuning the initial conditions.

There is one more important consideration. Above, we assumed that field evolution is monotonous so that $V(\tau)$ and $\phi(\tau)$ define a single-valued function $V(\phi)$. If $\phi'$ changes sign, that is, if $z$ crosses zero like in the models studied in \cite{Briaud:2023eae, Karam:2023haj} and as considered above, this is no longer true: the same $\phi$ values may be crossed multiple times, with different values of $V$, leading to an apparent contradiction. Note that this limitation is well-known in the context of the so-called ``horizon-flow'' program~\cite{Vennin:2014xta, Chowdhury:2019otk}, similar to the potential reconstruction presented here. 

We conclude that a single-field potential $V(\phi)$ can always be constructed in cases where $z$ does not change its sign, but care must be taken when interpreting models where a sign change happens. Such models may still be useful for a qualitative study, and it may be possible to produce the desired $z''/z$ behaviour for single-field inflation with a non-canonical kinetic term or for a multi-field setup. We leave such considerations for future work.

\section{Conclusions}
\label{sec:concl}
Most inflationary models designed to produce PBHs involve an initial constant-roll attractor phase, followed by a transition to a USR phase which finally leads to a second CR phase. As a result, the power spectrum is scale-invariant at large scales, then, as $k$ increases, it develops a dip, undergoes steep growth, and ultimately transitions to damped oscillations. However, in certain models, the dip is absent, and the conditions under which it appears have remained poorly understood until now.

In this work, we presented a detailed analysis of the linear dynamics of scalar perturbations when features are present in the inflaton potential. This was done by means of the transfer-matrix formalism that relates modes in the initial CR phase to modes in the final CR phase. It relies on the existence of exact solutions of the mode equation in those two phases, and is analogous to the S-matrix program that relates initial and final states in quantum field theory. By examining the properties of the transfer matrix, we derived efficient methods to approximate it across different regimes. At linear order in perturbation theory, this approach fully characterises the transitory phase and its impact on the curvature power spectrum.

At small scales, damped oscillations arise from non-trivial sub-Hubble evolution around the time of the transition. If $\tau_0$ denotes the moment of the transition and $\Delta\tau$ its duration, these spectral oscillations proceed at frequency $2\tau_0$ and are damped as a power law at $k<\Delta\tau^{-1}$ and exponentially at $k>\Delta\tau^{-1}$. How tall the oscillations are thus depends on how sharp the transition is. We have illustrated this behaviour in various models where we confirmed our findings.

At large scales, the dip (when it exists) results from non-trivial gradient effects at mildly super-Hubble scales around the time of the transition. We have established that a necessary and sufficient condition for the existence of a dip is that $z$ does not flip sign as it evolves. Since $z\propto \dot{\phi}$, $z$ flips sign when the inflaton turns around and starts rolling backwards in its potential.

We estimated the power spectrum at the dip and found a universal relationship between its value and the value of the power spectrum at the peak: the power spectrum at the peak scales as the inverse square of its value at the dip. This connects physics at two different scales and opens up new questions. First, the result appears as a general feature of single-field models of inflation and only relies on the assumption of a high peak (though it was shown to hold numerically even for moderate peaks). 
Whether or not this applies to multi-field scenarios is an open question -- if not, it offers an interesting way to distinguish between models. Second, it implies that PBH models that feature a high peak also contain a deep dip. It has been argued that the presence of a dip can have astrophysical consequences~\cite{Ozsoy:2021pws, Balaji:2022zur} although these still need to be investigated. Nevertheless, any structures (or lack thereof) appearing at the scales of the dip would be correlated with PBHs. Therefore, in models where the inflaton does not turn around, PBHs cannot exist without them. Third, corrections to our result can arise when going beyond the leading order in perturbation theory. At the classical level, the non-linear evolution has been addressed via lattice simulations~\cite{Caravano:2024tlp, Caravano:2024moy}, which show a decent agreement with analytic perturbative methods. Additionally, the perturbative power spectrum receives contributions from quantum corrections~\cite{Senatore:2009cf}. Although their effect on the peak is expected to be small, they can affect the dip by reducing its depth~\cite{Franciolini:2023agm, Cheng:2023ikq}.
Non-perturbative methods such as the stochastic-$\delta N$ formalism may also provide further insight into the possible smearing of the dip under quantum diffusion~\cite{Vennin:2015hra, Ando:2020fjm}.

\acknowledgments

This work was supported by the Estonian Research Council grants PSG869, PRG1055, PRG1677, PSG761, RVTT3 and RVTT7 and the Center of Excellence program TK202. E.T. was supported by the Lancaster--Manchester--Sheffield Consortium for Fundamental Physics under STFC grant: ST/T001038/1.

\appendix

\section{Regularising the mode functions at large  scales}
\label{app:Hadamard}
In the $k \ll \mathcal{H}$ super-Hubble limit, the Mukhanov--Sasaki mode functions follow the forms $u^{(0)}_{1,\mathrm{sup}}$ and $u^{(0)}_{2,\mathrm{sup}}$ given in \Eq{eq:u_0}. The second form,
\be
    u^{(0)}_{2,\mathrm{sup}}(\tau) = z(\tau) \, \int^{\tau}_{0} \frac{\td \tau'}{z^2(\tau')} \, ,
\ee
can become singular when $z$ crosses $0$ and must thus be regularised. In this appendix, we show that the correct regularisation scheme to be employed is Hadamard regularization.

We consider a scenario where $z$ vanishes at a time $\tau_*$, around which the Mukhanov--Sasaki frequency $z''/z$ remains finite and smooth. This implies that the mode functions $u^{(0)}_{1,\mathrm{sup}}$ and $u^{(0)}_{2,\mathrm{sup}}$ remain finite and smooth too. From the Wronskian condition, $u^{(0)}_{1,\mathrm{sup}} u^{(0)\prime}_{2,\mathrm{sup}}-u^{(0)\prime}_{1,\mathrm{sup}}u^{(0)}_{2,\mathrm{sup}}=1$, one must have $u^{(0)}_{2,\mathrm{sup}}(\tau_*)=-1/z'(\tau_*)$, since $u^{(0)}_{1,\mathrm{sup}}=z$ vanishes at $\tau_*$. As a consequence, $z'(\tau_*)\neq 0$. Moreover, since $z(\tau_*)=0$ but $z''/z$ remains finite at $\tau_*$, one must have $z''(\tau_*)=0$ and we can write
\bea
\label{eq:z:exp:tau0}
z(\tau) = z'(\tau_*) (\tau-\tau_*) + \mathcal{O}(\tau-\tau_*)^3
\eea 
around the singular point. 

In general, the second mode function can be written in the piecewise manner
\bea
\label{eq:u2_piecwise}
u^{(0)}_{2,\mathrm{sup}}(\tau) =
\begin{cases}
    z(\tau)\int^{\tau}_{\tau_{\mathrm{A}}} \frac{\td \tau'}{z^{2}(\tau')}, \quad 
    & \tau \leq \tau_*\\
    z(\tau)\left[ \int^{\tau}_{\tau_B} \frac{\td \tau'}{z^{2}(\tau')} +  C \right], \quad
    & \tau > \tau_*
\end{cases} \, .
\eea
By setting $\tau_A < \tau_* < \tau_B$, one ensures that the integral term is finite at $\tau<\tau_*$ and $\tau>\tau_*$, although it may diverge at $\tau_*$. The constant $C$ is set by the requirement that $u^{(0)\prime}_{2,\mathrm{sup}}$ is continuous at $\tau_*$. Indeed, the derivative of the mode function
\be
    u^{(0)}_{2,\mathrm{sup}}{}'(\tau)
    = 
\begin{cases}
    \frac{1}{z(\tau)} + z'(\tau)\int^{\tau}_{\tau_{\mathrm{A}}} \frac{\td \tau'}{z^{2}(\tau')}, \quad 
    & \tau \leq \tau_*\\
    \frac{1}{z(\tau)} + z'(\tau)\left[ \int^{\tau}_{\tau_B} \frac{\td \tau'}{z^{2}(\tau')} +  C \right], \quad
    & \tau > \tau_*
\end{cases}
\ee 
is explicitly finite even at $\tau_*$: this is because the divergence contained in the $1/z$ term cancels out the one contained in the integral term, as can be checked explicitly by using \Eq{eq:z:exp:tau0} around $\tau_*$. Choosing $C$ such that $u^{(0)\prime}_{2,\mathrm{sup}}$ is continuous gives
\bea
C=\lim_{\epsilon\to 0} \left[\int_{\tau_A}^{\tau_*-\epsilon}\frac{\mathrm{d}\tau'}{z^2(\tau')} - \int_{\tau_B}^{\tau_*+\epsilon}\frac{\mathrm{d}\tau'}{z^2(\tau')} +\frac{1}{z'(\tau_*)z(\tau_*-\epsilon)} - \frac{1}{z'(\tau_*)z(\tau_*+\epsilon)}\right] .
\eea 
As a consequence, after the singular point one has
\bea \label{eq:hadamard}
    u^{(0)}_{2,\mathrm{sup}}(\tau > \tau_*) 
    =& z(\tau) \lim_{\eps\to 0}\Bigg[ \int^{\tau}_{\tau_B} \frac{\td \tau'}{z^{2}(\tau')}     \\
    & + \int^{\tau_* -\eps}_{\tau_{\mathrm{A}}} \frac{\td \tau'}{z^{2}(\tau')} 
    + \int^{\tau_B}_{\tau_* +\eps} \frac{\td \tau'}{z^{2}(\tau')}
    + \frac{1}{z'(\tau_*) z(\tau_* - \eps)} - \frac{1}{z'(\tau_*) z(\tau_* + \eps)}  \Bigg] \\
    \equiv & z(\tau) \, \underline{\mathcal{H}}\!\!\int^{\tau}_{\tau_A} \frac{\td \tau'}{z^{2}(\tau')}\,,
\eea
where we recognise Hadamard's integral of finite parts, denoted with $\underline{\mathcal{H}}$ (not to be confused with the conformal Hubble parameter $\mathcal{H}$).

Note that Hadamard regularisation may change the sign of an integral. For instance, if $t_1<0<t_2$, then one has $\underline{\mathcal{H}}\int_{t_1}^{t_2} \mathrm{d}t/t^2 = 1/t_1-1/t_2<0$, although the integrand is always positive. This means that the regulatisation scheme has important consequences for the sign of the correction arising from the decaying mode.

To be more specific, during our second CR phase, we can write without loss of generality
\bea
\label{eq:z second CR phase}
z(\tau) = \beta\left(-\tau\right)^{1/2-\nu_\II}\left[1\pm\left(\tau/\tau^*\right)^{2\nu_\II}\right]\mbox{,} \qquad \tau \ge \tau_\II\, .
\eea
With the minus sign, $z$ crosses zero and flips sign at time $\tau^* > \tau_{\II}$. With the plus sign, $z$ never crosses zero. Using \Eqs{eq:z second CR phase} and \eqref{eq:hadamard}, we can explicitly compute the Hadamard integral $I$ needed for $A_2$ in \Eq{eq:A2 approximation}:
\be \label{eq:I_integrated}
I=\int^{\tau_\II}_{\tau_\I} \frac{\td \tau'}{z(\tau')^2} - \frac{\tau_\II}{2\nu_\II z\left(\tau_\II\right)^2}\left[1\pm\left(\frac{\tau_\II}{\tau^*}\right)^{2\nu_\II}\right] \, .
\ee
In the limit of strong enhancement (a long USR phase), we expect $|\tau^*| \ll |\tau_\II|$ and then, depending on whether $z$ flips sign of not, we obtain
\be
I= \mp\frac{\tau_\II}{2\nu_\II z\left(\tau_\II\right)^2}\left(\frac{\tau_\II}{\tau^*}\right)^{2\nu_\II} \, .
\ee
Thus, if the spectrum is strongly enhanced, the sign of $A_2$ only depends on whether $z$ flips sign or not, as discussed in \Sec{sec:dip_crit}.

\section{$A_4$ in the limit of strong power-spectrum enhancement}
\label{app:A4}

In this appendix, we show that in the limit of strong enhancement, the coefficient $A_4$ introduced in \Eq{eq:Pzeta expansion small k} diverges like $I=\underline{\mathcal{H}}\left[\int^0_{\tau_\I} \frac{\td \tau'}{z(\tau')^2}\right]$ and not like $I^2$ as one might naively expect. This has the important consequence that in this limit, $A_2^2\gg A_4$, because $A_2$ also scales as $I$, and $I$ is large.

First, using \Eq{eq:Different coefficients expansion small k}, one can show that, in the limit of strong enhancement, that is if $z$ approaches zero at a time $\tau^*$, $A_4$ should be dominated by
\be
\label{eq:dominant contribution A4}
\frac{A_4}{\kone^4} \approx -\frac{\tau_\I^2u^{(0)}_{1,\mathrm{sup}}(\tau_\I)^2u^{(0)}_{2,\mathrm{sup}}(\tau_\I)^2}{2} - \int_{\tau_\I}^0\td \tau'u^{(0)}_{2,\mathrm{sup}}(\tau')^2\int_{-\infty}^{\tau'}\td \tau''u^{(0)}_{1,\mathrm{sup}}(\tau'')^2 + \frac{1}{2\kone^4}A_2^2 \, .
\ee
All these different terms scale as $I^2$, which is why one might expect that $A_4$ also scales as $I^2$. However, the right-hand side of \Eq{eq:dominant contribution A4} turns out to vanish in the limit of strong enhancement. First, by integration by parts, the second term yields
\be
\begin{split}
    \int_{\tau_\I}^0\td \tau'u^{(0)}_{2,\mathrm{sup}}(\tau')^2\int_{-\infty}^{\tau'}\td \tau''u^{(0)}_{1,\mathrm{sup}}(\tau'')^2 =& -\frac{\tau_\I^2u^{(0)}_{1,\mathrm{sup}}(\tau_\I)^2u^{(0)}_{2,\mathrm{sup}}(\tau_\I)^2}{2} \\
    &- \int_{\tau_\I}^0\frac{\td \tau'}{z(\tau')^2}\int_0^{\tau'}\frac{\td \tau''}{z(\tau'')^2}\left[\int_{-\infty}^{\tau'}\td \tau''u^{(0)}_{1,\mathrm{sup}}(\tau'')^2\right]^2 \, .
\end{split}
\ee
Noticing that the integral in the right-hand side is dominated by the value of the integrand at $\tau^*$ where $z$ becomes small, we obtain
\be
\frac{A_4}{\kone^4} \approx \frac{1}{2\kone^4}A_2^2 -\frac{K^2I^2}{2} \, ,
\ee
where $K = \int_{-\infty}^{\tau^*}\td \tau'z(\tau')^2$. Then, using \Eq{eq:A2 approximation}, we obtain that at order $I^2$, the right-hand side of the previous equation vanishes, hence $A_4$ scales at most as $I$. We can thus conclude that $A_2^2\gg A_4$ when the power spectrum is strongly enhanced.

This last result can be explicitly checked in the case of an instantaneous transition (see \Sec{sec:instantaneous_transitions}). Indeed, in that model, the coefficient $A_4$ can be computed exactly and the result is
\be
\begin{split}
    \frac{A_4}{\kone^4} =& \frac{\tau_0^4}{16\left(\nu_\II+1\right)\left(\nu_\II+2\right)\left(\nu_\I-1\right)\left(\nu_\I-2\right)} \\
    &\left\lbrace\frac{\left(\nu_\I-1\right)\left(\nu_\I-2\right)}{2} + \left(\nu_\II+1\right)\left(\nu_\II+2\right) - \left(\nu_\I-2\right)\left(\nu_\II+2\right) 
    \right. \\ & \left.
    - \frac{\zeta_1}{\zeta_2\nu_\II}\left[\left(\nu_\II+1\right)\left(\nu_\II+2\right) - \left(\nu_\I-1\right)\left(\nu_\I-2\right)\right]\right\rbrace \, .
\end{split}
\ee
The limit of strong enhancement in this model corresponds to $\zeta_1\gg\zeta_2$. We thus find that $A_4$ scales linearly with the ratio $\zeta_1/\zeta_2$, that is, in the same way as $A_2$ [see \Eq{eq:A2 instantaneous}] and not as the square of this ratio. This confirms the conclusion that $A_2^2\gg A_4$.

\section{Useful formulae for the numerical computation of $A_2$, $A_4$, and $B_0$}
\label{app:Numerical formulae}

In \Eq{eq:Different coefficients expansion small k}, we obtained an expression for the coefficients $A_2$, $A_4$, and $B_0$ in terms of integrals between an initial time $\tau_\I$, corresponding to the beginning of the feature, and future infinity at $\tau=0$. However, $z$ diverges at future infinity, so in practice, the integrals must be computed until a finite time $\tau_\II$. Besides, $z$ can vanish in the second CR stage, so some integrals appearing in \eqref{eq:Different coefficients expansion small k} must be regularized, see Appendix~\ref{app:Hadamard}. In this appendix, we give expressions for the coefficients $A_2$, $A_4$ and $B_0$ that can be easily computed numerically. Taking $\tau_\II$ to be the end of the feature in the potential (the beginning of the second constant-$\nu$ phase, during  $\II_{\mr{USR}}$), and using the expression for $z$ in this phase from \Eq{eq:z second CR phase}, one is led to
\be
\begin{split}
\frac{A_2}{\kone^2} = &\frac{\tau_\I^2}{4\left(\nu_\I-1\right)} + \tau_\I z(\tau_\I)^2 \int_{\tau_\I}^{\tau_\II}\frac{\td \tau'}{z^2(\tau')} + \int_{\tau_\I}^{\tau_\II}\td\tau'z(\tau')^2\int_{\tau_\II}^{\tau'}\frac{\td \tau''}{z^2(\tau'')} \\
&-\frac{\tau_\II^2}{4\nu_\II}\left[1\pm\frac{\left(\tau_\II/\tau^*\right)^{2\nu_\II}}{1+\nu_\II}\right] \\
&\mp \frac{\left(-\tau^*\right)^{2\nu_\II}}{2\nu_\II\beta^2}\left[1-\frac{1}{1\pm\left(\frac{\tau_\II}{\tau^*}\right)^{2\nu_\II}}\right]\left[\int_{\tau_\I}^{\tau_\II}\td\tau'z(\tau')^2-\tau_\I z(\tau_\I)^2\right] \, ,
\end{split}
\ee

\be
\begin{split}
\frac{B_0}{\kone^{2\nu_\I}} =& -\left(\frac{-\tau_\I}{2}\right)^{2\nu_\I}\frac{\Gamma\left(1-\nu_\I\right)}{\tau_\I\Gamma\left(1+\nu_\I\right)} \\
&\left\lbrace 2\nu_\I z(\tau_\I)^2\left[\int_{\tau_\I}^{\tau_\II}\frac{\td \tau'}{z^2(\tau')} \pm \frac{\left(-\tau^*\right)^{2\nu_\II}}{2\nu_\II\beta^2}\left(1-\frac{1}{1\pm\left(\frac{\tau_\II}{\tau^*}\right)^{2\nu_\II}}\right)\right]+\tau_\I\right\rbrace \, ,
\end{split}
\ee
and
\be
\begin{split}
    \frac{A_4}{\kone^4} = & \frac{1}{2\kone^4}A_2^2 -\frac{\left(-\tau_\II\right)^4}{32\nu_\II^2\left(\nu_\II+1\right)}\left[1\mp \frac{4}{2+\nu_\II}\left(\frac{\tau_\II}{\tau^*}\right)^{2\nu_\II}+\frac{\left(\frac{\tau_\II}{\tau^*}\right)^{4\nu_\II}}{\nu_\II+1}\right] \\
    &+ \frac{1}{8\left(\nu_\I-1\right)^2}\left[-\frac{\tau_\I^4}{4\left(2-\nu_\I\right)} -\tau_\I^2u^{(0)}_{1,\mathrm{sup}}(\tau_\I)^2u^{(0)}_{2,\mathrm{sup}}(\tau_\I)^2 + \frac{\tau_\I^3u^{(0)}_{1,\mathrm{sup}}(\tau_\I)u^{(0)}_{2,\mathrm{sup}}(\tau_\I)}{2-\nu_\I}\right] \\
    &+ \tau_\I u^{(0)}_{1,\mathrm{sup}}(\tau_\I)^2\left(\left\lbrace\mp \frac{\left(-\tau^*\right)^{2\nu_\II}}{2\nu_\II\beta^2}\left[1-\frac{1}{1\pm\left(\frac{\tau_\II}{\tau^*}\right)^{2\nu_\II}}\right]\right\rbrace^2 \int_{\tau_\I}^{\tau_\II}\td\tau'z(\tau')^2 \right. \\
    &\left. + 2\left\lbrace\mp \frac{\left(-\tau^*\right)^{2\nu_\II}}{2\nu_\II\beta^2}\left[1-\frac{1}{1\pm\left(\frac{\tau_\II}{\tau^*}\right)^{2\nu_\II}}\right]\right\rbrace\int_{\tau_\I}^{\tau_\II}\td\tau'z(\tau')^2\int_{\tau_\II}^{\tau'}\frac{\td \tau''}{z^2(\tau'')} \right. \\
    &\left. + \int_{\tau_\I}^{\tau_\II}\td\tau'z(\tau')^2\left[\int_{\tau_\II}^{\tau'}\frac{\td \tau}{z^2(\tau)}\right]^2 + \frac{\left(-\tau_\II\right)^{2\nu_\II+2}}{8\beta^2\nu_\II^2\left(\nu_\II+1\right)}\right) \\
    &- \left\lbrace\mp \frac{\left(-\tau^*\right)^{2\nu_\II}}{2\nu_\II\beta^2}\left[1-\frac{1}{1\pm\left(\frac{\tau_\II}{\tau^*}\right)^{2\nu_\II}}\right]\right\rbrace^2\int_{\tau_\I}^{\tau_\II}\td\tau'z(\tau')^2\int_{\tau_\I}^{\tau'}\td \tau''z^2(\tau'') \\
    &- 2\left\lbrace\mp \frac{\left(-\tau^*\right)^{2\nu_\II}}{2\nu_\II\beta^2}\left[1-\frac{1}{1\pm\left(\frac{\tau_\II}{\tau^*}\right)^{2\nu_\II}}\right]\right\rbrace \\
&\times\int_{\tau_\I}^{\tau_\II}\td\tau'z(\tau')^2\left[\int_{\tau_\I}^{\tau'}\td \tau''z^2(\tau'')\right]\left[\int_{\tau_\II}^{\tau'}\frac{\td \tau}{z^2(\tau)}\right] \\
    &- \int_{\tau_\I}^{\tau_\II}\td\tau'z(\tau')^2\left[\int_{\tau_\I}^{\tau'}\td \tau''z^2(\tau'')\right]\left[\int_{\tau_\II}^{\tau'}\frac{\td \tau}{z^2(\tau)}\right]^2 \\
    &- \frac{\left(-\tau_\II\right)^{2\nu_\II+2}}{8\beta^2\nu_\II^2\left(\nu_\II+1\right)}\int_{\tau_\I}^{\tau_\II}\td\tau'z(\tau')^2 \, .
\end{split}
\ee

\bibliography{PBHinflation}

\end{document}